\documentclass{aa}
\usepackage{graphicx}
\usepackage{adjustbox}
\usepackage{tabularx}
\usepackage{xtab,booktabs}
\usepackage{txfonts}
\usepackage{natbib}
\usepackage{caption}
\usepackage[colorlinks=true,linkcolor=black,urlcolor=blue,citecolor=blue]{hyperref} 

\begin{document}

\title{The CARMENES search for exoplanets around M dwarfs}  
\subtitle{A deep transfer learning method to determine $T_{\rm eff}$ and [M/H] of target stars}

\titlerunning{A deep transfer learning method for fundamental parameters of M dwarfs}
\author{A.~Bello-Garc\'ia\inst{\ref{ovi}}
        \and
        V.\,M.~Passegger\inst{\ref{iac},\ref{uiac},\ref{hs},\ref{ou}}
        \and
        J.~Ordieres-Mer\'e\inst{\ref{upm}}
        \and
        A.~Schweitzer\inst{\ref{hs}}
            \and
        J.\,A.~Caballero\inst{\ref{cab}}
        \and
        A.~Gonz\'alez-Marcos\inst{\ref{ur}}
        \and
        I.~Ribas\inst{\ref{ice},\ref{ieec}}
        \and
        A.~Reiners\inst{\ref{iag}}
        \and
        A.~Quirrenbach\inst{\ref{lsw}}
        \and
        P.\,J.~Amado\inst{\ref{iaa}}
        \and
        V.\,J.\,S.~B\'ejar\inst{\ref{iac},\ref{uiac}}
        \and 
        C.~Cifuentes\inst{\ref{cab}}
        \and
        Th.~Henning\inst{\ref{mpia}}
        \and
        A.~Kaminski\inst{\ref{lsw}}
        \and
        R.~Luque\inst{\ref{iaa},\ref{uch}}
        \and
        D.~Montes\inst{\ref{ucm}}
        \and
        J.\,C.~Morales\inst{\ref{ice},\ref{ieec}}
        \and
        S.~Pedraz\inst{\ref{caha}}
        \and 
        H.\,M.~Tabernero\inst{\ref{cab1}}
        \and
        M.~Zechmeister\inst{\ref{iag}}
        }

\institute{
        Departamento de Construcci\'on e Ingenier\'ia de Fabricaci\'on, Universidad de Oviedo, c/ Pedro Puig Adam, Sede Departamental Oeste, M\'odulo 7, 1$^a$ planta, E-33203 Gij\'on, Spain \newline
        \email{abello@uniovi.es}\label{ovi}
        \and
        Instituto de Astrof\'{\i}sica de Canarias, c/ V\'ia L\'actea s/n, 38205 La Laguna,
        Tenerife, Spain \label{iac}
        \and
        Departamento de Astrof\'{\i}sica, Universidad de La Laguna, 38206 La Laguna, Tenerife, Spain \label{uiac}
        \and
        Hamburger Sternwarte, Gojenbergsweg 112, D-21029 Hamburg, Germany \label{hs}
        \and
        Homer L. Dodge Department of Physics and Astronomy, University of Oklahoma, 440 West Brooks Street, Norman, OK 73019, United States of America \label{ou}
        \and
        Departamento de Ingenier\'ia de Organizaci\'on, Administraci\'on de Empresas y Estad\'istica, Universidad Polit\'ecnica de Madrid, c/ Jos\'e Guti\'errez Abascal 2, E-28006 Madrid, Spain \label{upm}
            \and
        Centro de Astrobiolog\'ia (CSIC-INTA), ESAC, Camino bajo del castillo s/n, E-28692 Villanueva de la Ca\~nada, Madrid, Spain \label{cab}
        \and
        Departamento de Ingenier\'ia Mec\'anica, Universidad de la Rioja, c/ San Jos\'e de Calasanz 31, 26004 Logro\~no, La Rioja, Spain \label{ur}
        \and
        Institut de Ci\`encies de l'Espai (CSIC-IEEC), Campus UAB, c/ de Can Magrans s/n, 08193 Bellaterra, Barcelona, Spain \label{ice}
        \and
        Institut d'Estudis Espacials de Catalunya (IEEC), 08034 Barcelona, Spain \label{ieec}
        \and
        Institut f\"ur Astrophysik und Geophysik, Georg-August-Universit\"at, Friedrich-Hund-Platz 1, 37077 G\"ottingen, Germany \label{iag}
        \and
        Landessternwarte, Zentrum f\"ur Astronomie der Universit\"at Heidelberg, K\"onigstuhl 12, 69117 Heidelberg, Germany\label{lsw}
        \and
        Instituto de Astrof\'isica de Andaluc\'ia (IAA-CSIC), Glorieta de la Astronom\'ia s/n, 18008 Granada, Spain \label{iaa}
        \and
        Max-Planck-Institut f\"ur Astronomie, K\"onigstuhl 17, 69117 Heidelberg, Germany \label{mpia}
        \and
        Department of Astronomy and Astrophysics, University of Chicago, Chicago, IL 60637, United States of America \label{uch}
        \and
        Departamento de F\'{\i}sica de la Tierra y Astrof\'{\i}sica and IPARCOS-UCM (Instituto de F\'{\i}sica de Part\'{\i}culas y del Cosmos de la UCM), Facultad de Ciencias F\'{\i}sicas, Universidad Complutense de Madrid, 28040 Madrid, Spain \label{ucm}
        \and
        Centro Astron\'omico Hispano en Andaluc\'ia (CAHA), Observatorio de Calar Alto, Sierra de los Filabres, 04550 G\'ergal, Almer\'ia, Spain\label{caha}
        \and
        Centro de Astrobiolog\'ia (CSIC-INTA), Carretera de Ajalvir km 4, Torrej\'{o}n de Ardoz, 28850, Madrid, Spain \label{cab1}
}

\date{Received 3 May 2022 / Accepted 22 March 2023}

\abstract
{The large amounts of astrophysical data being provided by existing and future instrumentation require efficient and fast analysis tools. 
 Transfer learning is a new technique promising higher accuracy in the derived data products, with information from one domain being transferred to improve the accuracy of a neural network model in another domain. 
In this work, we demonstrate the feasibility of applying the deep transfer learning (DTL) approach to high-resolution spectra in the framework of photospheric stellar parameter determination. To this end, we used 14 stars of the CARMENES survey sample with interferometric angular diameters to calculate the effective temperature, as well as six M dwarfs that are common proper motion companions to FGK-type primaries with known metallicity.
After training a deep learning (DL) neural network model on synthetic PHOENIX-ACES spectra, we used the internal feature representations together with those 14+6 stars with independent parameter measurements as a new input for the transfer process.
We compare the derived stellar parameters of a small sample of M dwarfs kept out of the training phase with results from other methods in the literature. Assuming that temperatures from bolometric luminosities and interferometric radii and metallicities from FGK+M binaries are sufficiently accurate, 
DTL provides a higher accuracy than our previous state-of-the-art DL method (mean absolute differences improve by 20~K for temperature and 0.2~dex for metallicity from DL to DTL when compared with reference values from interferometry and FGK+M binaries). Furthermore, the machine learning (internal) precision of DTL also improves as uncertainties are five times smaller on average.
These results indicate that DTL is a robust tool for obtaining M-dwarf stellar parameters comparable to those obtained from independent estimations for well-known stars.}
%
\maketitle
%

\section{Introduction}
\label{introduction}

The determination of photospheric stellar parameters in M dwarfs has always been challenging. 
M dwarfs are smaller, cooler, and fainter than Sun-like stars. 
Because of their faintness and their higher stellar activity, with sometimes stronger magnetic fields, stronger line blends, and the lack of true continuum, well-established photometric and spectroscopic methods are brought to their limits. 
In the literature, there are several methods to estimate M-dwarf photospheric parameters, such as effective temperature ($T_{\rm eff}$), surface gravity ($\log{g}$), and metallicity ([M/H]); for example, spectroscopic indices \citep[see][]{RojasAyala2012,GaidosMann2014}, photometric relations \citep[see][]{Dittmann2016,Houdebine2019}, interferometry \citep[see][]{Boyajian2012,vonBraun2014,Rabus2019}, synthetic model fits \citep[see][]{Gaidos2014,Passegger2018,marfil2021carmenes}, and machine learning \citep[ML; see][]{Antoniadis2020,Passegger2020}.

One method considered to be relatively precise is calibration with M dwarfs that have a late F, G, or early K common proper motion companion with known metallicity. 
Many of the relations mentioned below were calibrated using FGK+M multiple systems \citep[e.g.,][]{Newton2014}. As a representative example, \cite{Mann2013a} identified spectral features sensitive to metallicity in low-resolution optical and near-infrared (NIR) spectra of 112 late-K to mid-M dwarfs in multiple systems with earlier companions, from which they derived different metallicity calibrations. 
The same relations were used by \cite{Rodriguez2019} to determine metallicity from mid-resolution $K$-band spectra for 35 M dwarfs of the {\it K2} mission. 
Other photometric calibrations using FGK+M binary systems were presented by \cite{Bonfils2005}, \cite{Casagrande2008}, \cite{JohnsonApps2009}, \cite{SchlaufmanLaughlin2010}, and \cite{Neves2012}, among others, while several spectroscopic calibrations were explored by \cite{Rojas-Ayala2010}, \cite{Dhital2012}, \cite{Terrien2012}, \cite{Mann2014}, \cite{Mann2015}, and, more recently, \cite{Montes2018}.

\begin{figure}
    \centering
    \includegraphics[width=0.49\textwidth]{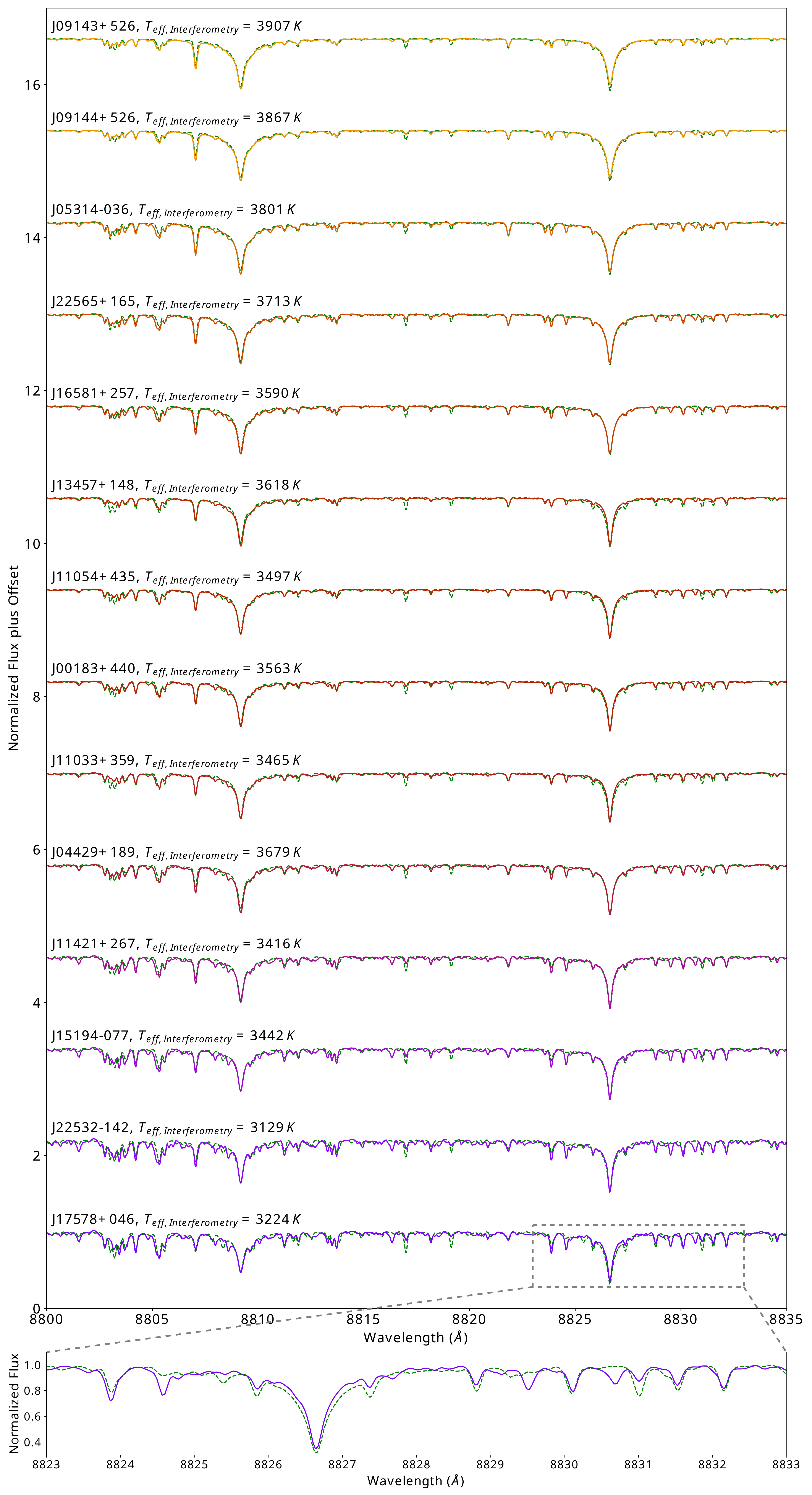}
    \caption{Observed vs. synthetic fluxes (8800--8835\,\r{A}) for stars with interferometric $T_{\rm eff}$.
    Solid rainbow: CARMENES normalized template spectra.
    Dashed green: PHOENIX normalized synthetic spectra.
    Stars are sorted by decreasing $T_{\rm eff}$
    from \cite{Schweitzer2019}, with interferometrically derived $T_{\rm eff}$ indicated above them. 
    {\it Bottom panel}: Zoom-in detail of one representative spectrum.}
    \label{fig:spectra_interf}
\end{figure}

\begin{figure}
    \centering
    \includegraphics[width=0.49\textwidth]{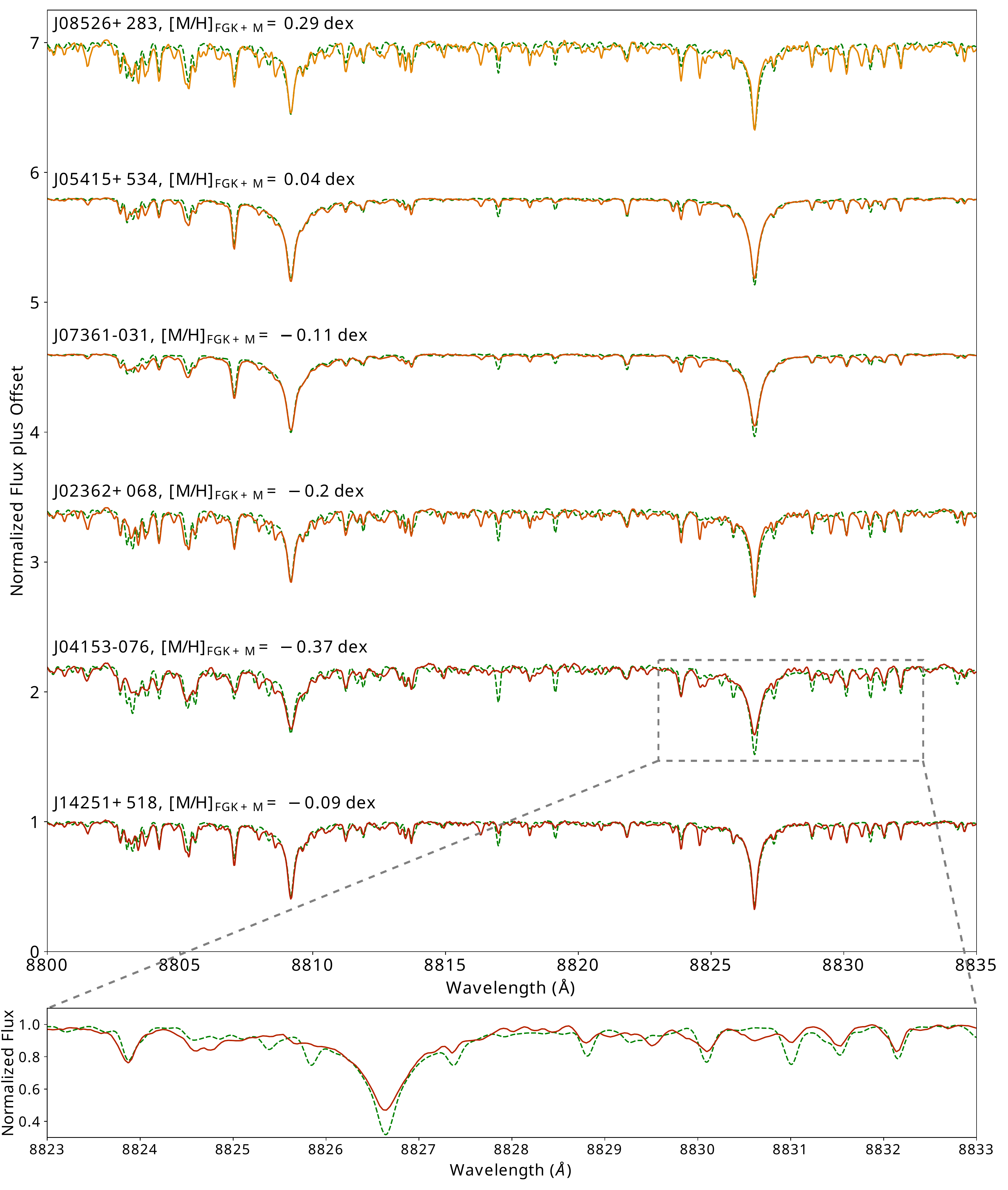}
    \caption{Same as Fig.~\ref{fig:spectra_interf} but for M-dwarf common proper motion companions to FGK stars with well-determined metallicity, and sorted by decreasing metallicity from \cite{Schweitzer2019}.
    }
    \label{fig:spectra_binaries}
\end{figure}

Fundamental stellar parameters can also be derived from interferometric measurements. 
However, only a limited number of late-type dwarfs are accessible for such observations because they must be bright and nearby.
\cite{Boyajian2012} presented interferometric angular diameters for 26 K and M dwarfs measured with the CHARA array and for 7 K and M dwarfs from the literature. 
With parallaxes and bolometric fluxes, these authors computed the absolute luminosity ($L$), radii ($R$), and $T_{\rm eff}$. 
They also calculated empirical relations for K0 to M4 dwarfs to connect $T_{\rm eff}$, $R$, and $L$ to a broadband color index and iron abundance [Fe/H].
On the other hand, \cite{Maldonado2015} estimated $T_{\rm eff}$ from pseudo-equivalent widths (pEWs) of temperature-sensitive lines calibrated with interferometric $T_{\rm eff}$ from \cite{Boyajian2012} and metallicities from pEWs calibrated with the relations of \citet{Neves2012}.
 \cite{Maldonado2015} constructed a mass--radius relation using interferometric radii \citep{Boyajian2012,vonBraun2014} and masses from eclipsing binaries \citep{Hartman2015}. 
From this, they calculated $\log{g}$. 
Other {studies} that derived M-dwarf $T_{\rm eff}$ from angular diameters include, for example, those of \cite{Segransan2003}, \cite{Demory2009}, \cite{vonBraun2014}, and \cite{Newton2015}. 
Of these, \cite{Segransan2003} also determined $\log{g}$ from their measured masses and radii. 

\begin{figure*}
  \centering
  \includegraphics[width=0.48\textwidth]{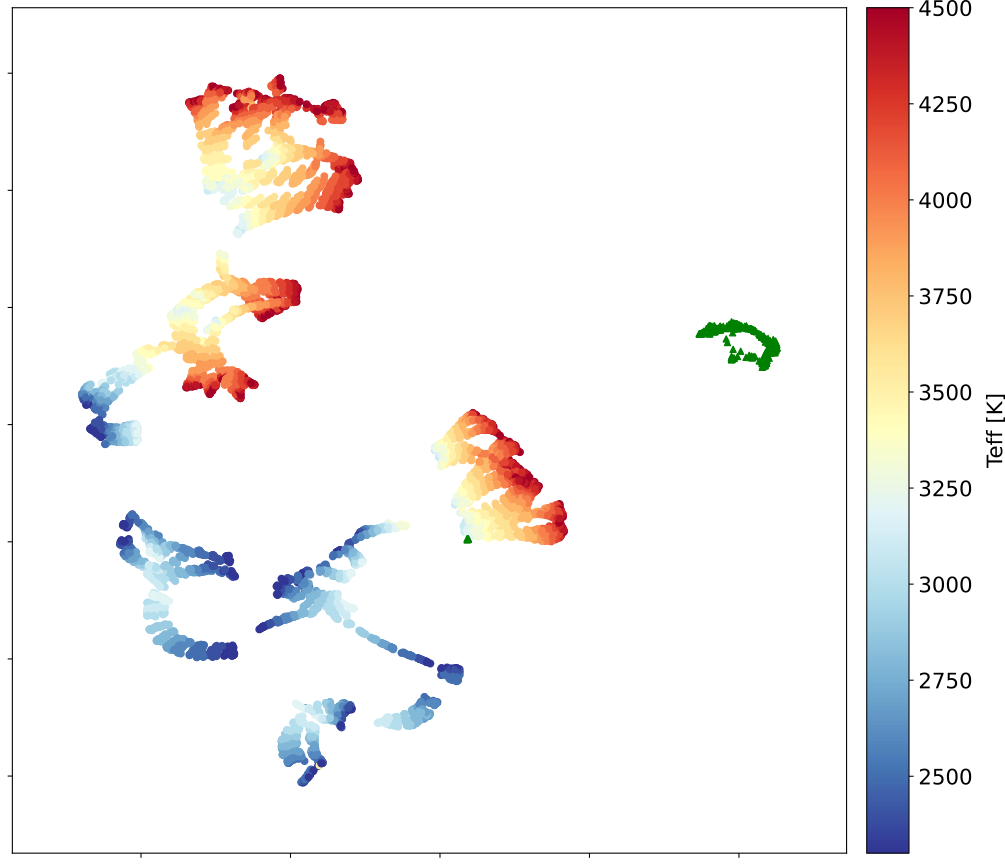}
  \hspace{0.5cm}
  \includegraphics[width=0.48\textwidth]{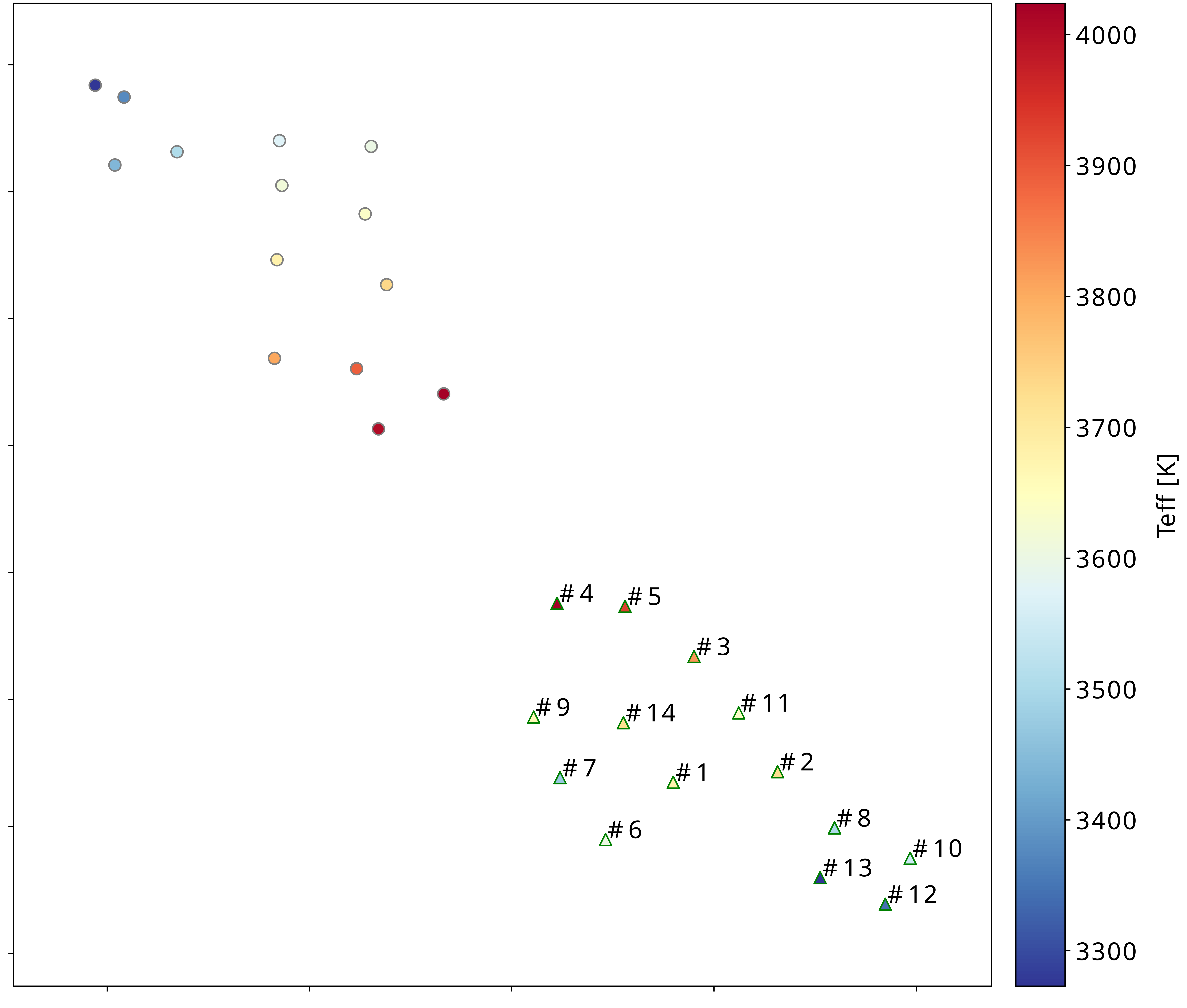}  
  \caption{Representative two-dimensional UMAP projections of observed and synthetic spectra {from the 8800--8835\,\AA{} window, with $T_{\rm eff}$ color coded.}
  {\it Left panel:} PHOENIX-ACES set used for DL training  and  the 282 CARMENES spectra (green).
  {\it Right panel:} Subsample of 14 CARMENES M dwarfs with interferometric $T_{\rm eff}$ values shown in Table~\ref{tab:interf_teff} (green triangles and labeled). Colored circles represent their closest interpolated best-fit PHOENIX model, using \cite{Schweitzer2019} parameter estimations as a reference.}
  \label{fig:umap_teff_phx_carmenes}
\end{figure*}

Different approaches have been taken to estimate photospheric stellar parameters for M dwarfs in general, mainly within the paradigm of comparing measured {line fluxes} with theoretical ones calculated {from} different sets of synthetic spectra.
Although different algorithms using $\chi^2$-minimization or principal component analysis have been {employed}, several artificial intelligence techniques have also been proposed.
For these, the differences between observation and theory are not at the level of individual lines, but {are based on whole spectral} regions~\citep{fabbro2018application,kielty2018starnet,bialek2020assessing,minglei2020atmospheric,Passegger2020}. 
Indeed, some comparisons of techniques regarding the estimation of stellar parameters have already been carried out~\citep{Passegger2022}.
However, there are still several open questions related to the uncertainty of parameter estimation {due to} the signal-to-noise ratio (S/N) of the flux signal, and {due to} the {``synthetic gap''} \citep{Fabbro2018,Tabernero2022}, which is the difference in feature distribution between theoretical and observed spectra.
The impact of the synthetic gap can be appreciated in Figs.~\ref{fig:spectra_interf} and~\ref{fig:spectra_binaries}, where we display a specific flux window showing one Mg~{\sc i} line at 8809\,\AA{} and one Fe~{\sc i} line at 8827\,\AA{} in the red optical for two sets of M dwarfs. 
Even the very latest synthetic models show slight differences with respect to high-S/N, high-resolution CARMENES\footnote{Calar Alto high-Resolution search for M dwarfs with Exo-earths with Near-infrared and optical \'Echelle Spectrographs, \url{https://carmenes.caha.es}.} spectra, especially with faint lines for which parameters are not yet well constrained.
In Figs.~\ref{fig:spectra_interf} and~\ref{fig:spectra_binaries}, the sequence of spectra is ordered according to $T_{\rm eff}$ and metallicity from \citet{Schweitzer2019}, which are not identical to the parameters estimated from interferometry or binary companions, as shown in the figures.
This mismatch is an effect caused by the synthetic gap. Another effect can be seen by the flux differences of observed and synthetic spectra in the bottom panels of Figs.~\ref{fig:spectra_interf} and~\ref{fig:spectra_binaries} (zoomed-in spectra).

\citet{o2020interpreting} presented an interesting demonstration of the spectra transfer process, although their focus was rather different than ours. These authors showed that transferred spectra can reduce the synthetic gap from the pure physical models, which is further evidence of the value of transfer technologies.

The spatial dimension of features (i.e., the number of flux points within the wavelength window) depends on the size of the flux range, but it is usually very high (e.g., 3500 dimensions in the case of Figs.~\ref{fig:spectra_interf} and \ref{fig:spectra_binaries}). Therefore, some specific techniques are needed to project a spectrum from such high-dimensional space into a lower one, while preserving inter-distances that help to better understand the topology. 
To this end,~\citet{Passegger2020} introduced a technique to visualize the relative positions of a set of spectra in the 2D Euclidean space, the uniform manifold approximation and projection \citep[UMAP; ][]{mcinnes2018umap-software}. 
The main purpose of these projections is to illustrate the difference in feature distribution between synthetic and observed spectra, that is, the so-called synthetic gap.
In the left panel of Fig.~\ref{fig:umap_teff_phx_carmenes}, different theoretical PHOENIX spectra are projected along with high-S/N, high-resolution, telluric-subtracted spectra observed with CARMENES. 
To make the synthetic spectra comparable to the observed ones, before plotting we included continuum normalization and instrumental and rotational broadening. 
However, no noise was added, as it was shown in Fig.~4 of \cite{Passegger2020} that adding only noise has a negligible effect on the projection.
In this representation, no stellar parameters are involved and the UMAP only depends on the flux values of every spectrum.
The theoretical feature map only partially covers the CARMENES range (green circles), with a significant part of the spectra projected far away from them. 
Some patterns emerge when additional information, such {as} a color code for $T_{\rm eff}$, is incorporated into the UMAP, as illustrated in both panels of Fig.~\ref{fig:umap_teff_phx_carmenes}, {independent of} the source of $T_{\rm eff}$: theoretical (left) or interferometric (right).

 In this work, to reduce the uncertainties associated {with} the synthetic gap and therefore enable {a more reliable estimation of stellar parameters}, we propose a way to bridge the synthetic gap and transfer the knowledge from measured flux signals estimated by interferometry and FGK+M systems to the features derived from the theoretical models used with deep learning (DL). 
Such an approach is known as deep transfer learning (DTL)~\citep{tan2018survey, awang2020classification, wei2020deep}. 
For $T_{\rm eff}$, we transfer knowledge gained from interferometrically determined $T_{\rm eff}$ for a few stars to the rest of the CARMENES spectra, while for metallicity we transfer knowledge gained from spectral synthesis of FGK stars.
However, this DTL technique requires a significant amount of data, which is problematic because of the limited number of high-resolution spectra for stars fulfilling those conditions. 
This is even worse when data-based modeling techniques are used, as they require a methodology to assess the quality of the created model when applied to stars not used during the training phase. 
Despite these limitations, we show that the proposed technique is valid and its accuracy will increase as more stars with independent estimates of their parameters are incorporated.

In this paper, we use DTL to determine new $T_{\rm eff}$ and [M/H] for 286 M dwarfs from the 
CARMENES survey \citep{Reiners2018a,Quirrenbach2020}, and compare our results with the literature. As our technique is based on our previous work on DL, we refer to \cite{Passegger2020} for further information.
The basic workflow of the DTL can be summarized as follows: (1) train DL models on a large set of synthetic model spectra, (2) extract the internal feature representations (3) train DTL models based on the external knowledge about stellar parameters that was transferred to the neural network, (4) calculate stellar parameter estimations for the stars.
In Sect.~\ref{Methods} we explain the DTL procedure and our artificial neural network (ANN) architecture. 
Section~\ref{Analysis} describes the values obtained from the literature (interferometry and FGK+M systems) for each parameter that we used for training the ANN, the stellar sample, and the application of our ANN. 
The derived stellar parameters are presented in Sect.~\ref{Results}, together with a literature comparison and discussion. 
Finally, in Sect.~\ref{Summary} we provide a short summary. 


\begin{table*}
\caption{Interferometrically derived $T_{\rm eff}$ values transferred. }
\label{tab:interf_teff}
\centering 
\tiny 
\setlength{\tabcolsep}{0.4em}
\begin{tabular}{llllr@{\,$\pm$\,}lcr@{\,$\pm$\,}lcr@{\,$\pm$\,}lr@{\,$\pm$\,}lcr@{\,$\pm$\,}l}
    \hline 
    \hline 
    \noalign{\smallskip}
 \# & Karmn      & Gliese   &      Name       &\multicolumn{2}{c}{$L$}& Ref. $L$  &\multicolumn{2}{c}{$d$}& Ref. $d$& \multicolumn{2}{c}{$S^{(a)}$}   & \multicolumn{2}{c}{$\theta_{\rm LD}$} & Ref. $\theta_{\rm LD}$ & \multicolumn{2}{c}{$T_{\rm eff}$} \\
   &         &          &                 &\multicolumn{2}{c}{[$10^{-4}L_\sun$]}&      & \multicolumn{2}{c}{[pc]}&     &\multicolumn{2}{c}{[$10^{-11}$W m$^{-2}$]}& \multicolumn{2}{c}{[mas]} &                    & \multicolumn{2}{c}{[K]} \\
    \noalign{\smallskip}
    \hline 
    \noalign{\smallskip}
 1 & J00183+440 &  15A &         GX\,And &  239.1 &  9.2 & This work &   3.56244 &   0.00026 & EDR3 &  6.03 & 0.23 & 1.005 & 0.005 & Boy12 & 3658   & 36 \\
 2 & J04429+189 &  176 &      HD\,285968 &  358.1 &  8.9 &     Cif20 &   9.4730 &   0.0063 &  DR2 &  1.28 & 0.03 & 0.448 & 0.021 & vBr14 & 3717   & 90 \\
 3 & J05314--036 &  205 &       HD\,36395 &  657.0 &  3.9 & This work &   5.70408 &   0.00066 & EDR3 &  6.46 & 0.04 & 0.943 & 0.004 & Boy12 & 3843 & 10 \\
 4 & J09143+526 & 338A &       HD\,79210 &  811   & 30   &     Cif20 &   6.3339 &   0.0015 &  DR2 &  6.47 & 0.24 & 0.834 & 0.014 & Boy12 & 4087   & 51 \\
 5 & J09144+526 & 338B &       HD\,79211 &  759   & 15   &     Cif20 &   6.3337 &   0.0017 &  DR2 &  6.05 & 0.12 & 0.856 & 0.016 & Boy12 & 3968   & 42 \\
 6 & J11033+359 &  411 &  Lalande\,21185 &  225.2 &  8.9 & This work &   2.54613 &   0.00021 & EDR3 & 11.11 & 0.44 & 1.432 & 0.013 & Boy12 & 3571   & 39 \\
 7 & J11054+435 & 412A &    BD+44\,2051A &  193.4 &  4.3 &     Cif20 &   4.848 &   0.024 & HIP2 &  2.63 & 0.06 & 0.764 & 0.017 & Boy12 & 3411   & 43 \\
 8 &J11421+267 &  436 &       Ross\,905 &  243.1 &  3.4 &     Cif20 &   9.7560 &   0.0089 &  DR2 &  0.82 & 0.01 & 0.417 & 0.013 & vBr12 & 3446   & 55 \\
 9 & J13457+148 &  526 &      HD\,119850 &  377   & 13   &     Cif20 &   5.4353 &   0.0015 &  DR2 &  4.09 & 0.14 & 0.835 & 0.014 & Boy12 & 3642   & 43 \\
 10 & J15194--077 &  581 &         HO\,Lib &  123.5 &  2.6 &     Cif20 &   6.2992 &   0.0021 &  DR2 &  1.00 & 0.02 & 0.446 & 0.014 & vBr11 & 3501   & 58 \\
 11 & J16581+257 &  649 &     BD+25\,3173 &  447   & 20   &     Cif20 &  10.3827 &   0.0034 &  DR2 &  1.33 & 0.06 & 0.484 & 0.012 & vBr14 & 3610   & 61 \\
 12 & J17578+046 &  699 & Barnard's\,star &  35.23 &  0.91 &     Cif20 &   1.82665 &   0.00097 &  DR2 &  3.38 & 0.09 & 0.952 & 0.005 & Boy12 & 3252   & 23 \\
 13 & J22532--142 &  876 &         IL\,Aqr &  127.1 &  2.5 &     Cif20 &   4.6758 &   0.0017 &  DR2 &  1.86 & 0.04 & 0.746 & 0.009 & vBr14 & 3165   & 25 \\
 14 & J22565+165 &  880 &      HD\,216899 &  516.8 &  8.9 &     Cif20 &   6.8676 &   0.0018 &  DR2 &  3.51 & 0.06 & 0.744 & 0.004 & Boy12 & 3713   & 19 \\
    \noalign{\smallskip}
    \hline
\end{tabular}%
\tablebib{
HIP2: \citet{HIP};
vBr11: \citet{von2011astrophysical};
Boy12: \citet{Boyajian2012};
vBr12: \citet{von2012gj};
vBr14: \citet{vonBraun2014};
DR2: \citet{DR2};
Cif20: \citet{cifuentes2020carmenes};
EDR3: \citet{EDR3}.
}
\tablefoot{
\tablefoottext{a}{ $S$ is the flux at Earth calculated from the tabulated values $L$ and $d$. $T_{\rm eff}$ is calculated using Eq.~\ref{eq:teff}.}
}
\end{table*}

\section{Methods}
\label{Methods}

The aim of ML is to automatically discover rules that must be followed in order to efficiently map input data to a desired output. In this process, it is essential to create appropriate representations of the data. These representations are task-dependent and may vary according to the final task that the selected ML algorithm is going to perform.
DL is a subfield of ML where a hierarchical representation of the data is created, and has received increasing attention in recent years in light of its successful application to numerous real-world problems (e.g., virtual assistants, visual recognition, fraud detection, machine translation, medical image analysis, photo descriptions; see \citealt{Karpathy2015} and many others).
The higher levels of the hierarchy are formed by the composition of representations of the lower level~\citep{Passegger2020}. 
More importantly, this hierarchy of representations is automatically learned from the data by completely automating the most crucial step in ML, namely feature engineering. 
Automatically learning features at multiple levels of abstraction allows a system to learn complex representations mapping the input to the output directly from the data, without completely depending on human-crafted features.
The word ``deep'' refers to the multiple hidden layers used to obtain those representations. 
In this sense, DL can also be called hierarchical feature engineering~\citep{sarkar2018hands}.

Data dependence is one of the most serious issues in DL, which is extremely dependent on massive training data {sets} when compared to traditional ML methods.
Although the amount of data needed depends on the type of model, the required accuracy, and the complexity of the model, all these factors can lead to the requirement for large datasets.
Therefore, an intrinsic and unavoidable problem has always been insufficient training data. Data collection is complex and expensive, making the generation of large-scale, high-quality annotated data sets extremely difficult. Therefore, techniques to work with data sets of limited size are of great value, as is the case for the DTL technique.

\subsection{Deep transfer learning} 
\label{sect:deep.transfer.learning}

It has become increasingly common in various domains, such as image recognition and natural language processing, to pre-train the entire model in a data-rich task \citep{kraus2017decision,gao2018deep,raffel2019exploring,HAN2021225}. 
Ideally, this pre-training process causes the model to develop general-purpose abilities and knowledge that can then be transferred to downstream tasks. 
\citet{Goodfellow-et-al-2016} referred to {transferred learning} (TL) in the context of generalization. 
These latter authors defined TL as the situation where what has been learned in one setting is exploited to improve generalization in another setting. Therefore, TL provides a robust and practical solution to leverage information from one domain to improve the accuracy of a model built for a different domain \citep{Vilalta2018}.

\citet{5288526} proposed a more precise definition of TL, starting by defining a domain and a task, respectively. A domain can be represented by $ D = {\chi, P (X)} $, which contains two parts: the feature space $\chi$ and the marginal probability distribution $ P (X)$, where $ X = \{x_1, ..., x_n\} \in \chi $ . The task can be represented by $ T = \{y, f (x)\} $, and consists of two parts: a label space $ y $ and a target-prediction function $f(x)$. This function $ f (x) $ can also be regarded as a conditional probability function $P(y|x)$. Then, given a learning task $T_t$ based on $D_t$ (where the subscript $t$ refers to ``transferred''), TL is designed to improve the performance of a predictive function $f_T(\cdot)$ in learning the task $T_t$ by discovering and transferring latent knowledge from another domain $D_s$ and learning task $T_s$ (where the subscript $s$ refers to ``source'', which in our case is the PHOENIX-ACES synthetic models), where $D_s \neq D_t$ and/or $T_s \neq T_t$. Usually, the size of the source domain $D_s$ is much larger than the size of the transferred domain $D_t$ (i.e., $N_s \gg N_t$). Based on the previous definitions, a DTL task is defined by $<D_s,T_s,D_t,T_t,f_T(\cdot)>$, where $f_T(\cdot)$ is a nonlinear function involving a deep ANN. TL relaxes the hypothesis that the training data must be independent of and identically distributed with the test data, which motivates the use of TL for the problem of insufficient training data.

\begin{figure}
  \centering
  \includegraphics[width=0.49\textwidth]{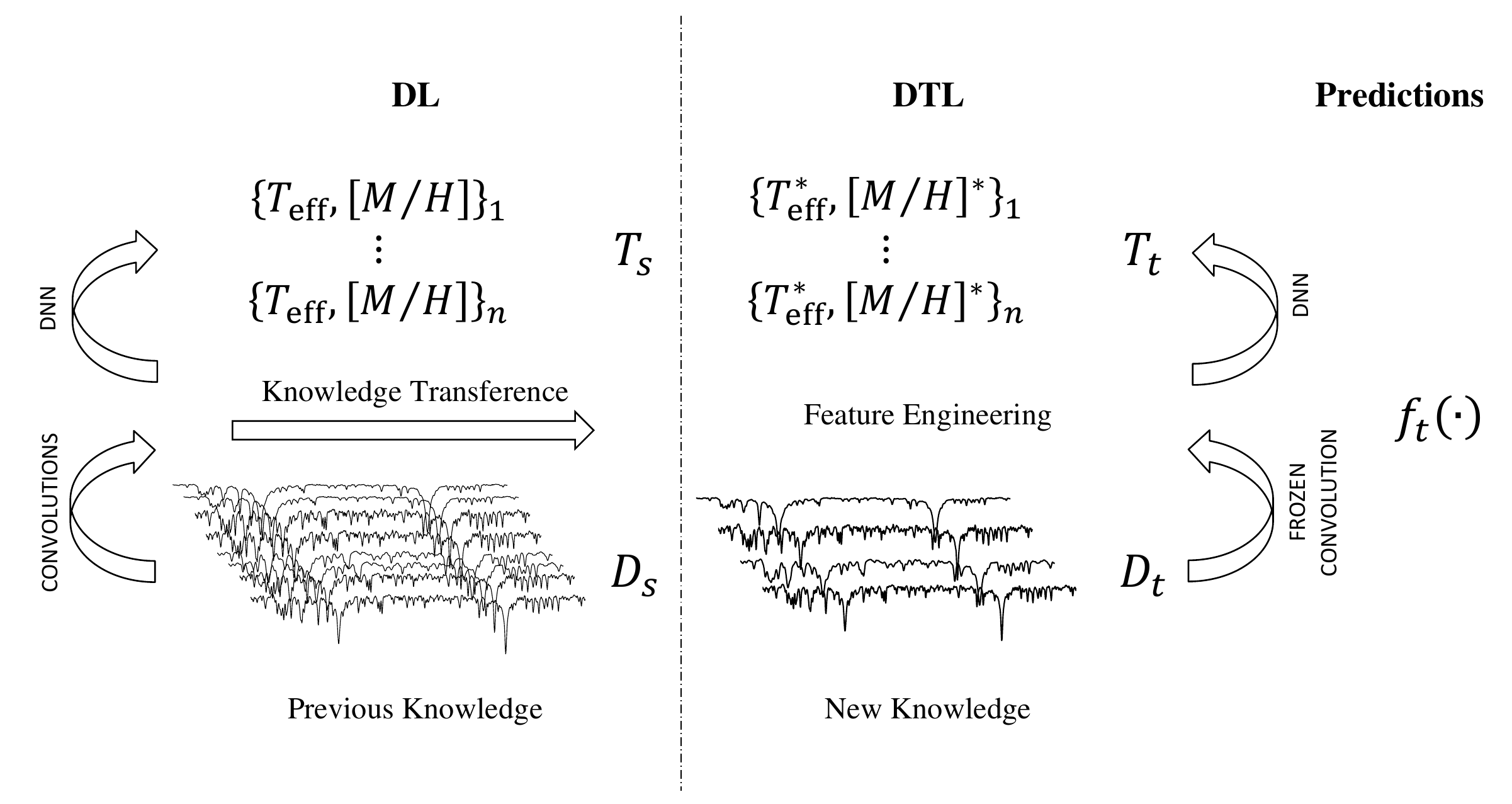}  
  \caption{Representation of the domains and tasks that the DTL process can be applied to.}
  \label{fig:domains}
\end{figure}

The popularity of DL has led to many different DTL methods, and several authors have proposed a classification of them~\citep{10.1007/978-3-030-01424-7_27,zhao2021applications}. Common categories involve instance-, mapping-, network-, and adversarial-based TL. Each of these categories has its particular applicability, considering the specific context and characteristics of domains and tasks $<D_{\rm s},T_{\rm s},D_{\rm t},T_{\rm t},f_{\rm T}(\cdot)>$. In our case, due to the characteristics of the problem, we selected the network category to implement our DTL.
The relationship between domains and tasks is illustrated by Fig.~\ref{fig:domains}.

\subsection{Artificial neural network architectures}

\begin{figure*}
  \centering
  \includegraphics[width=0.95\textwidth]{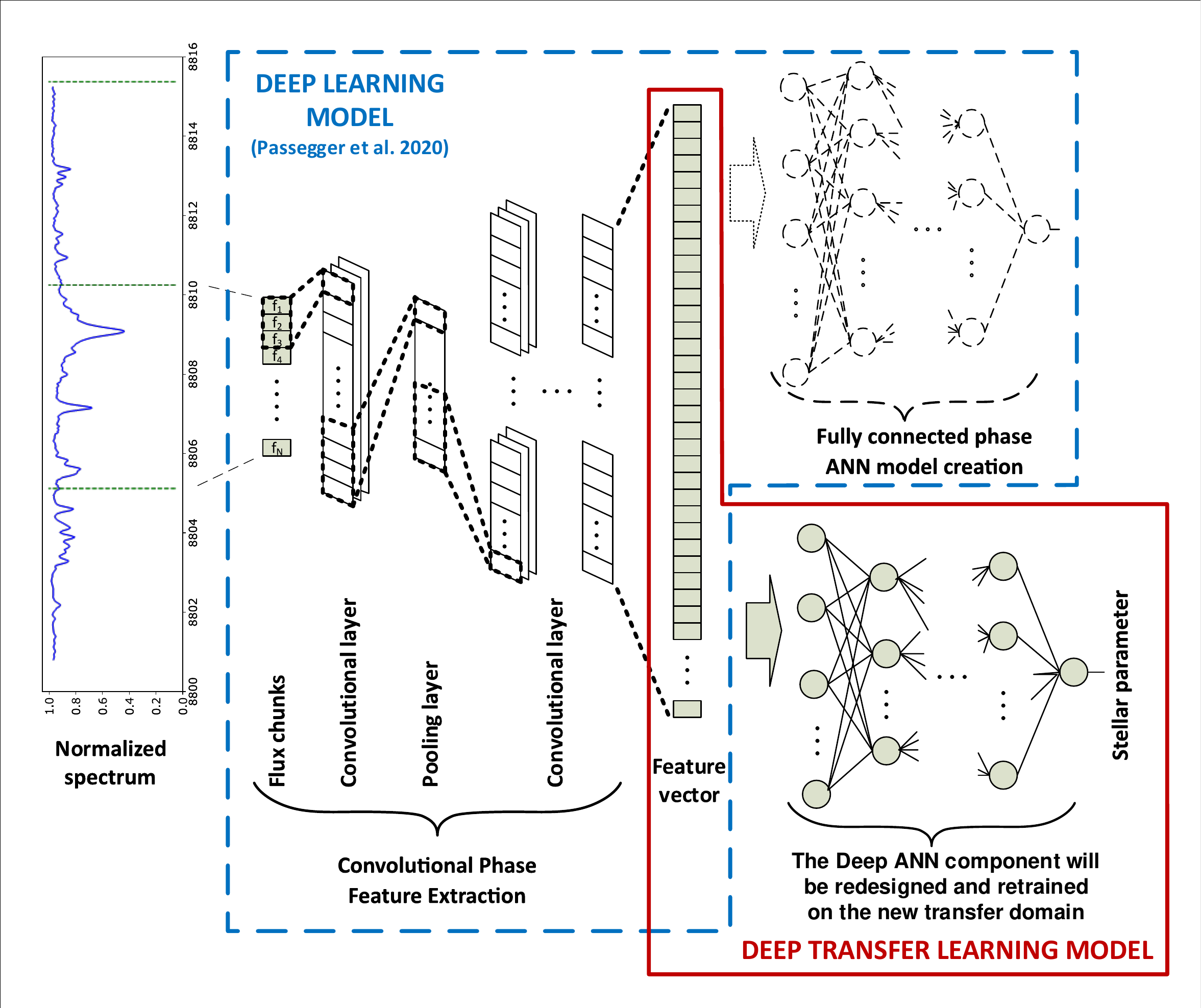}  
  \caption{Representation of the adopted strategy for the network-based approach, where the ANN component has been redesigned with respect to the DL model derived from PHOENIX-ACES spectra \citep[to be compared with Fig.~1 of][]{ Passegger2020}.}
  \label{fig:DTL}
\end{figure*}

Network-based DTL refers to reusing the part of the pre-trained network in the source domain, including its network structure and connection parameters, and transferring it to be a part of the deep neural network that is used in the target domain. 
The main assumption is connected with the idea that the features identified in the source domain will be valid in the transfer domain, whereas $f_T(\cdot)$ requires adaptation. 
As a result, we kept the original distribution of fully connected layers that we applied for the DL analysis by \cite{Passegger2020}. Moreover, \cite{Passegger2020} constructed several neural network models for different spectral regions, finding that the results for all regions are comparable, but that the region 8800--8835\,\AA{} gives the smallest validation error. Therefore, we also adopted this strategy and only use that region.
In this sense, we rely on the DL models already trained in \cite{Passegger2020} and extract the internal feature representations for each star in order to use it here as a new input for the DTL process. The DL models were trained on a grid of 449 806 synthetic PHOENIX-ACES spectra, after removing unphysical stellar parameter combinations not corresponding to main sequence stars using the PARSEC v1.2S evolutionary models \citep{Bressan2012,Chen2014,Chen2015,Tang2014}. 
The feature representation was taken from the flattened layer of the DL model, as represented in~Fig.~\ref{fig:DTL}.

\subsection{DTL training and testing}
\label{sect:dtl_training_testing}

\begin{table}[]
\caption{Supplementing literature $T_{\rm eff}$ values transferred from\,\citet{Passegger2022}.}
\label{tab:lit_teff}
\centering %
\begin{tabular}{lllc@{\,$\pm$\,}l}
    \hline
    \hline
    \noalign{\smallskip}
    Karmn    & Gliese   & Name   & \multicolumn{2}{c}{$T_{\rm eff}$ [K]} \\
    \noalign{\smallskip}
    \hline
    \noalign{\smallskip}
 J00067--075 & 1002 & G\,158-27 &  2844 & 149 \\
 J07558+883 & 1101  & G\,253-6  &  3217 & 68  \\
 J10508+068 & 402   & Wolf\,358 &  3169 & 93  \\
 J13005+056 & 493.1 & Wolf\,461 &  3090 & 107 \\
 J23419+441 & 905   & HH\,And   &  3058 & 88  \\
    \noalign{\smallskip}
    \hline
\end{tabular}%
\end{table}

To build the transfer domain $D_{\rm t}$ for $T_{\rm eff}$, we started with 14 stars of the CARMENES survey with interferometric angular diameters $\theta_{\rm LD}$ measured by \cite{Boyajian2012}, \cite{vonBraun2014}, and references therein.
We did not use the derived $T_{\rm eff}$ from these publications.
Instead, we used updated bolometric fluxes $S$ at Earth based on the most recent photometry (in particular, {\it Gaia} photometry for the optical passbands) collected by \citet{cifuentes2020carmenes} to calculate our reference $T_{\rm eff}$ with the distance-independent form of the Stefan-Boltzmann law:

\begin{equation}
\label{eq:teff}
T_{\rm eff} = \theta_{\rm LD}^{-1/2} \left(\frac{4 ~ S}{\sigma}\right)^{1/4}.
\end{equation}
As \citet{cifuentes2020carmenes} did not actually tabulate the bolometric flux at Earth $S$, we used the tabulated bolometric luminosities $L$ and the distances $d$, which were used in measuring $L$, and calculated $S$ via $S=L/(4\pi d^2)$.
All used and derived values ($L$, $d$, $S$, $\theta_{\rm LD}$, and $T_{\rm eff}$) are listed in Table~\ref{tab:interf_teff}.

As for Gl\,15A our derived $T_{\rm eff}$ from Eq.~\ref{eq:teff} was in severe disagreement with $T_{\rm eff}$ listed in \cite{Boyajian2012}, we realized that for this bright star, as well as for the bright star Gl\,411, the luminosities in \cite{cifuentes2020carmenes} were missing reliable J band photometry. Therefore, we revise the luminosity determinations of \cite{cifuentes2020carmenes} by adding classical Johnson J band photometry as well as the {\it Gaia} Early Data Release 3 (EDR3) data. Gl\,205 was not included in \cite{cifuentes2020carmenes}, but its $L$ is derived in the same fashion. These three stars are marked in Table~\ref{tab:interf_teff} with `This work'.

As the 14 stars of the interferometric sample include only two M dwarfs with $T_{\rm eff} < 3440$~K, we supplemented them with five mid-to-late M dwarfs listed in Table~\ref{tab:lit_teff}, for which a good $T_{\rm eff}$ estimation is available in the literature, as recommended by~\citet{Passegger2022}.
This was done in order to obtain training and validation sets with regularly spaced parameters. To achieve such regular distribution, we binned the to-be-transferred data set of 19 M dwarfs into a variable number of bins.
The goal was to have as many as 75\,\% nonempty bins. For each of those bins, we selected a representative element.
If two or more stars were included in one bin, we picked the closest to the midpoint of the bin.

\begin{table*}[]
\caption{Transferred spectroscopically determined [M/H] values from FGK+M systems.}
\label{tab:binaries_zeta}
\centering %
\begin{tabular}{lll r@{\,$\pm$\,}l lll}
    \hline 
    \hline 
    \noalign{\smallskip}
    Karmn     & Gliese   & Name           &  \multicolumn{2}{c}{{[}M/H{]}} & \multicolumn{2}{c}{FGK primary} &  Reference \\
     & & & \multicolumn{2}{c}{(dex)} & Name & Sp. type & \\
    \noalign{\smallskip}
    \hline 
    \noalign{\smallskip}
    J02362+068 & 105B  & BX Cet            & $-0.20$ & 0.02 & HD 16160 A & K0\,V+ & \citet{Montes2018} \\
    J04153--076 & 166C  & $o^{02}$ Eri C    & $-0.37$ & 0.02 & $o^{02}$ Eri A & K0.5\,V &  \citet{Montes2018}\\
    J05415+534 & 212   & HD 233153         & $+0.04$ & 0.02 & V538 Aur & K1\,V & \citet{Montes2018} \\
    J07361--031 & 282C  & BD--02 2198        & $-0.11$ & 0.03 & V869 Mon & K2\,V & \citet{Montes2018} \\
    J08526+283 & 324B  & $\rho^{01}$ Cnc B & $+0.29$ & 0.04 & $\rho^{01}$ Cnc A & G8\,V & \citet{Montes2018} \\
    J14251+518 & 549B  & $\theta$ Boo B    & $-0.09$ & 0.01 & $\theta$ Boo A & F7\,V & \cite{Tabernero2022} \\
    \noalign{\smallskip}
    \hline
\end{tabular}%
\end{table*}

\begin{table}[]
\caption{Supplementary literature [M/H] values transferred from\,\citet{Passegger2022}.}
\label{tab:lit_zeta}
\centering %
\begin{tabular}{lllr@{\,$\pm$\,}lc}
   \hline
    \hline
    \noalign{\smallskip}
    Karmn    & Gliese   & Name   & \multicolumn{2}{c}{[M/H]} \\
    \noalign{\smallskip}
    \hline
    \noalign{\smallskip}
J00067--075 & 1002  & G\,158-27       & $-0.19$ & 0.16 \\ 
J00183+440 & 15A   & GX And          & $-0.26$ & 0.09 \\
J04429+189 & 176   & HD\,285968      & $+0.08$ & 0.10 \\
J05314--036 & 205   & HD\,36395       & $+0.39$ & 0.11 \\
J07558+833 & 1101  & G\,253-6        & $+0.00$ & 0.10 \\
J09143+526 & 338A  & HD\,79210       & $-0.13$ & 0.12 \\
J09144+526 & 338B  & HD\,79211       & $-0.11$ & 0.14 \\
J10508+068 & 402   & EE\,Leo         & $+0.16$ & 0.11 \\
J11033+359 & 411   & Lalande\,21185  & $-0.31$ & 0.11 \\
J11054+435 & 412A  & BD+44\,2051A    & $-0.38$ & 0.12 \\
J11421+267 & 436   & Ross\,905       & $+0.00$ & 0.11 \\
J13005+056 & 493.1 & FN\,Vir         & $+0.09$ & 0.10 \\
J13457+148 & 526   & HD\,119850      & $-0.22$ & 0.11 \\
J15194--077 & 581   & HO\,Lib         & $-0.13$ & 0.11 \\
J16581+257 & 649   & BD+25\,3173     & $-0.04$ & 0.12 \\
J17578+046 & 699   & Barnard's star  & $-0.39$ & 0.12 \\
J22565+165 & 880   & HD\,216899      & $+0.18$ & 0.09 \\
J23419+441 & 905   & HH\,And         & $+0.25$ & 0.11 \\
    \noalign{\smallskip}
    \hline
\end{tabular}
\end{table}

For {the metallicity}, [M/H], the adopted strategy followed the same structure as for $T_{\rm eff}$. We reverted to metallicities measured for FGK stars with a proper motion M-dwarf companion. Due to their formation from the same cloud, it is assumed that both components share the same metallicity \citep{Desidera2006,Andrews2018}. Table~\ref{tab:binaries_zeta} presents five M dwarfs with CARMENES spectra with an FGK primary with known metallicity obtained from \cite{Montes2018}, and one from \cite{Tabernero2022}, namely J14251+518 (\object{$\theta$~Boo~B}).
Due to the low number of multiple systems, our transfer strategy is to use the list of 18 stars from \cite{Passegger2022} in Table~\ref{tab:lit_zeta} combined with those listed in Table~\ref{tab:binaries_zeta} as $D_{\rm t}$. Although the values from \cite{Passegger2022} are not as accurate as those from binaries, they {do not depend strongly on a specific model} because they were calculated as medians from several literature values.

To avoid the potential lack of generalization linked to accepting models based solely on their performance over the validation set, we propose using a more robust methodology, which is designed to create two different and separate groups of samples: one set for training and validation, and the rest of the stars as a test group to measure the quality of the models.
Indeed, because of the relatively significant influence that a single sample can have on the model performance, which is due to the low number of samples in the training or validation subsets, we used the cross-validation training approach. Cross-validation is a data-resampling method to assess the generalization ability of predictive models and to prevent overfitting~\citep{refaeilzadeh2009cross}. Briefly, the data are usually divided into two segments: one used for training a neural network model and one used for validating the trained model. The basic form of cross-validation is a $k$-fold cross-validation, where the data are divided into $k$ folds (in our case, we set $k=4$) before $k$ iterations of training and validation are conducted, where in each iteration a different fold is kept aside for validation, while the remaining $k-1$ folds are used for training.

DTL model creation requires a quality criterion to assess the learning progress of the ANN. The quality criterion that is often adopted is a threshold in the loss error during the validation process (the loss function is widely used in mathematical optimization and decision theory). Looking to obtain a sufficient variety of models due to randomness in the selection of samples and the optimization starting point, several repetitions of the model-creation process were accomplished.
As a four-fold cross-validation strategy was adopted, four potentially valid models were created per repetition of the model-creation process, and another four validation loss errors were measured at the end of the training processes. The trained models were deemed of sufficient quality when the validation error was lower than 0.01. In order to have a significant set of models in the case of metallicity, 80 repetitions of the model-creation process were adopted, which means $80 \times 4 = 320$ potential models. 
Only 121 of these reached convergence under the adopted criterion, and they were later used for predictions. In the case of $T_{\rm eff}$, 20 repetitions of the model-creation process were adopted and all the $20 \times 4 = 80$ potential models reached convergence. This behavior indirectly shows that, as expected, the $T_{\rm eff}$ parameter has more power than metallicity in terms of impact on the spectra.

It is straightforward to build a model that is perfectly adapted to the data set at hand but then unable to generalize to new and unseen data. 
Therefore, the value of measuring model quality over an independent data set becomes evident.
In this way, the model performance can be externally assessed using the estimation provided by the information gathered during the validation step~\citep{vabalas2019machine}.

\section{Analysis}
\label{Analysis}

\subsection{Observational sample}
\label{Analysis.Sample}

\begin{figure*}
    \centering
    \includegraphics[width=0.48\textwidth]{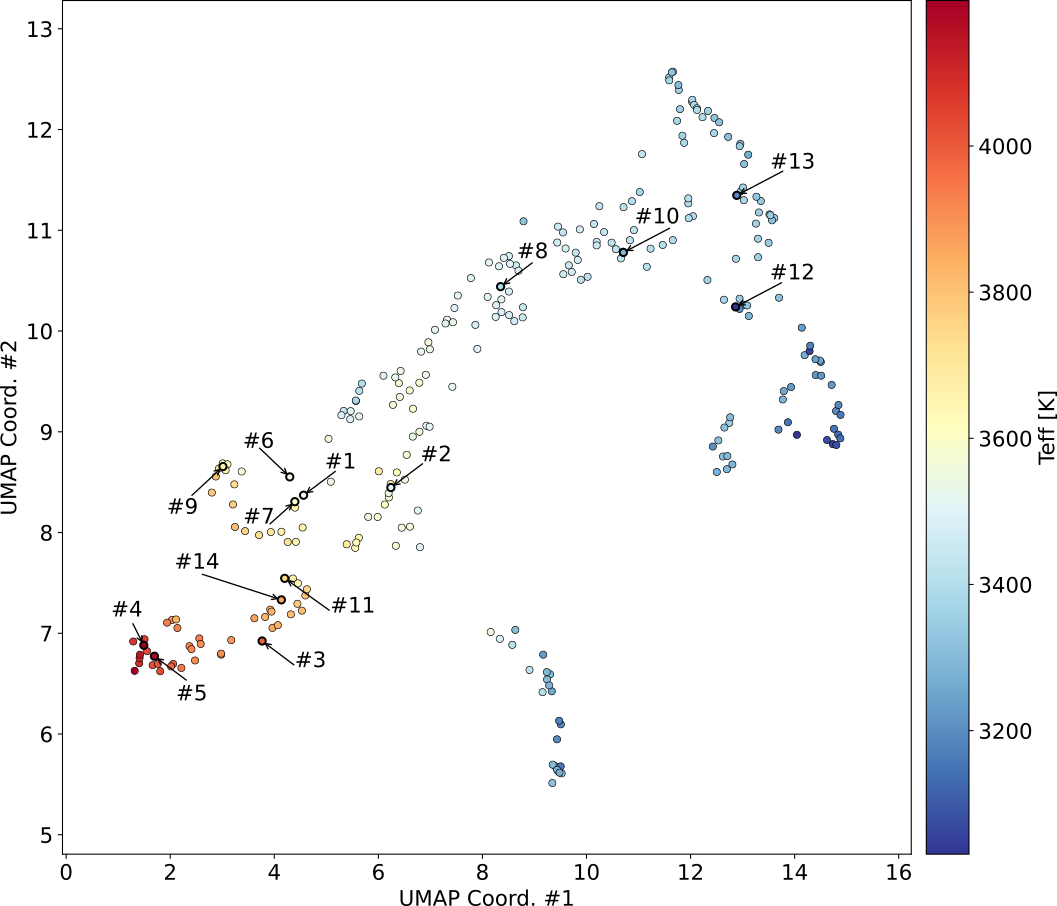}
    \hspace{0.5cm}
    \includegraphics[width=0.47\textwidth]{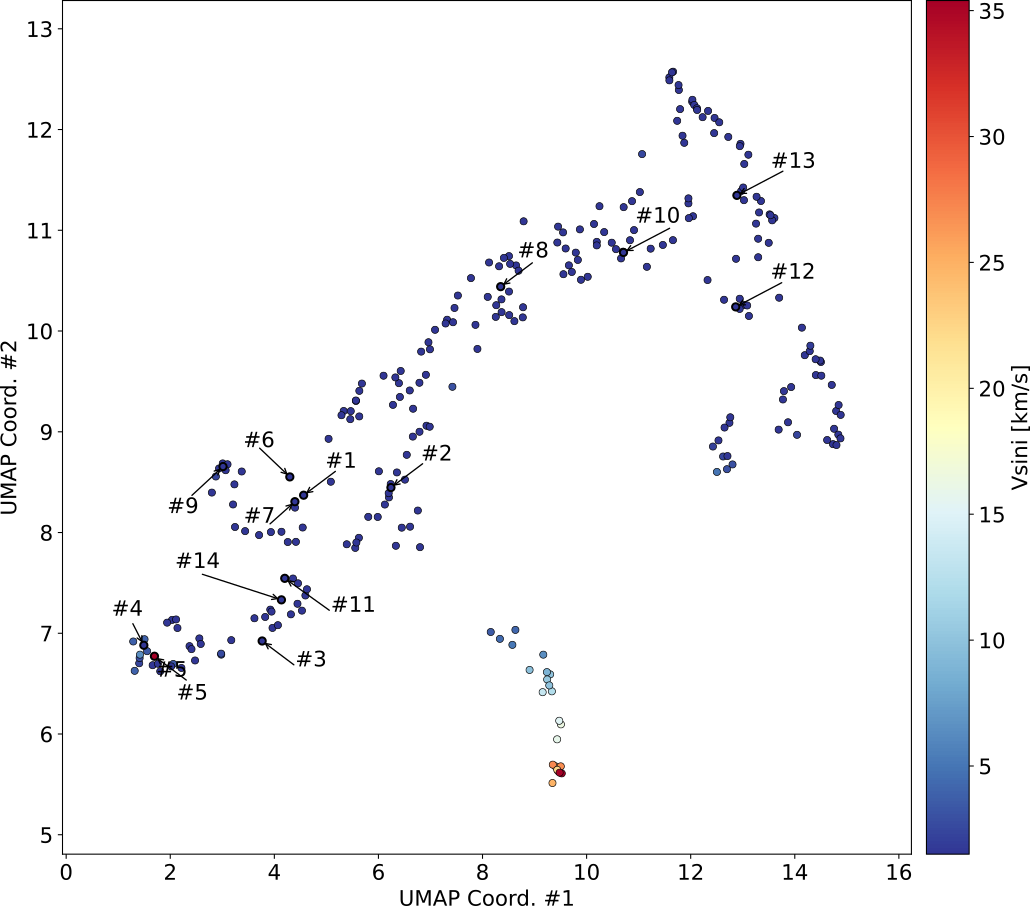}
    \caption{UMAP of CARMENES color-coded according to $T_{\rm eff}$ ({\it left}) and $\varv\sin{i}$ ({\it right}), built with the flux values from the 8800--8835\,{\AA} window. 
    The M dwarfs with interferometric values are labeled. 
    $T_{\rm eff}$ values
    in the left panel {are taken} from \cite{Schweitzer2019} and Table~\ref{tab:interf_teff},  while rotational velocities in the right panel {are taken} from \cite{Passegger2019}.}
    \label{fig:umap_teff_carmenes_interf}
\end{figure*}

To test our DTL method, we used the same template spectra as in \cite{Passegger2019} and applied it to all 282 M dwarfs listed in their Table~B.1 plus four more stars coming from the independent interferometric sample used for the learning process. 
We focus here on a small sample of the galactic stellar population, namely M dwarfs of spectral type M0 to M6. To verify whether or not our method is able to generalize beyond this parameter range, it is necessary to apply it to a much larger stellar sample, such as APOGEE or {\it Gaia}, which shall be part of a subsequent study.
The stars were observed with CARMENES on the Zeiss 3.5\,m telescope at the Observatorio de Calar Alto, Spain. 
CARMENES combines two highly stable fiber-fed spectrographs covering a spectral range from 520\,nm to 960\,nm in the optical (VIS) and from 960\,nm to 1710\,nm in the NIR, with spectral resolutions of $R \approx$ 94\,600 and 80\,400, respectively \citep{Quirrenbach2018,Reiners2018a}.
The primary goal of this instrument is to search for Earth-sized planets in the habitable zones of M dwarfs \citep[e.g.,][]{Zechmeister2019}.

For a detailed description of our data-reduction procedure, we refer to \cite{Zechmeister2014}, \cite{Caballero2016}, and \cite{Passegger2019}. 
As in the latter, we used the high-S/N template (co-added) spectrum for each star. These templates are a byproduct of the CARMENES radial-velocity pipeline {\tt serval} 
\citep[SpEctrum Radial Velocity AnaLyser;][]{Zechmeister2018}. In the standard data flow, the code constructs a template for every target star from at least five individual spectra to derive the radial velocities of a single spectrum by least-square fitting to the template. 
For our sample, the average S/N of the order, in which our investigated wavelength window of 8800--8835\,\AA{} lies in the beginning of the order, amounts to $258 \pm 158$.

Before creating the templates, the NIR spectra were corrected for telluric lines. We did not use the telluric correction for the VIS spectra because the telluric features are negligible in the investigated range. The telluric correction was explained in detail by \cite{Nagel2022}. 

For normalization of our spectra, we used the same method and routine as in \cite{Passegger2020}, the Gaussian Inflection Spline Interpolation Continuum ({\tt GISIC}\footnote{\url{https://pypi.org/project/GISIC/}}, developed by D.\,D.~Whitten and designed for spectra with strong molecular features). 
After the spectrum was smoothed with a Gaussian, and continuum points were selected, the pseudo-continuum was normalized with a cubic spline interpolation. 
We applied the same procedure to both observed and synthetic spectra within the spectral window 8800--8835\,\AA{} by adding 5\,\AA{} on each side to avoid possible edge effects.
The observed spectra have been corrected for radial velocity to match the rest frame of the synthetic spectra using the cross-correlation \citep[crosscorrRV from PyAstronomy,][]{Czesla2019} between a PHOENIX model spectrum and the observed spectrum. To obtain a universal wavelength grid, which is necessary for applying the DL method, the wavelength grid of the observed spectra was linearly interpolated with the grid of the synthetic spectra.


\subsection{Transferred knowledge}
\label{transferred.knowledge}
In our particular case, where the distance between domains is significant (see Fig.~\ref{fig:umap_teff_phx_carmenes}) and the sample density in $D_{\rm t}$ is limited (see Tables~\ref{tab:interf_teff} and \ref{tab:binaries_zeta}), the network-based approach was selected as the adequate DTL method.
In Fig.~\ref{fig:umap_teff_carmenes_interf}, the stars with interferometric $T_{\rm eff}$ are regularly distributed in $T_{\rm eff}$ along the whole CARMENES data set. This is needed and must be checked before transferring any knowledge from another study, as DL and DTL techniques are not very good at extrapolating information because of the limited stellar parameter range of our transferred and training sets.

In terms of the terminology introduced in Sect.~\ref{sect:deep.transfer.learning}, our $D_{\rm s}$ domain was built over the PHOENIX-ACES spectra library \citep{Husser2013} with a flux window of 35\,{\AA} between 8800\,{\AA} and 8835\,{\AA} {in the VIS channel} for consistency with previous work~\citep{Passegger2020, Passegger2022}. Furthermore, $T_{\rm s}$ is the DL model that minimizes the error on a test set of unused PHOENIX-ACES spectra. {In other words, the DL model selected for transferring the feature space is the best one of those trained on synthetic PHOENIX-ACES spectra.}
The definition of $D_{\rm t}$ for the two stellar parameters is introduced in Sect.~\ref{sect:deep.transfer.learning}.

From this, up to 80 different transference models were created. If their training process reaches convergence, they were selected as contributing to the prediction of the stellar parameters of the stars in the test set. Proposals from different transferred models are collected and integrated using the kernel density estimate (KDE) technique. This technique allows us to establish the most frequent value for the stellar parameter, but also its uncertainty, which {depends} on the star and the flux window. This KDE estimation can be seen as the predictive function $f_{\rm T}(\cdot)$.


\subsection{Implementation}

As already explained in Sect.~\ref{sect:deep.transfer.learning}, transfer learning is an approach in DL (and ML) where knowledge is transferred from one model to another. 
This means that a properly trained DL model is the first step, as presented in Fig.~\ref{fig:DTL}.
For the DTL network-oriented approach, we kept the features selected by the DL model, which means freezing the convolutional transformers and providing a new deep ANN configuration. This enabled us to train the weights of the connections for stellar parameters according to the interferometric measurements, binary companion estimations, or both. 
The architectures used for convolutional layers and the deep neural network are presented in Tables~\ref{tab:convDL} and~\ref{tab:nnDTL}, respectively.

\begin{table}[]
\caption{Convolutional architecture used in the DL model.}
\label{tab:convDL}
\centering %
\begin{tabular}{lccc}
   \hline
    \hline
    \noalign{\smallskip}
Operation & Input size & Kernel size & Number of variables \\
    \noalign{\smallskip}
\hline
    \noalign{\smallskip}
Input              & 3501         & 1          &   ...    \\
Conv 1D            & 3501         & 32         & 128   \\
MaxPool            & 1750         &    ...        &   ...    \\
Conv 1D            & 1750         & 16         & 1552  \\
MaxPool            & 875          &   ...         &    ...   \\
Conv 1D            & 875          & 8          & 392   \\
MaxPool            & 437          &   ...         &    ...   \\
Conv 1D            & 437          & 4          & 100   \\
MaxPool            & 218          &   ...         & ...      \\
Flatten            & 872          &   ...         &    ...   \\
    \noalign{\smallskip}
\hline
\end{tabular}
\end{table}

\begin{table}[]
\caption{Neural network architecture used in the DTL model.}
\label{tab:nnDTL}
\centering %

\begin{tabular}{lcc}
\hline
\hline
    \noalign{\smallskip}
Operation & Input size & Variables \\
    \noalign{\smallskip}
\hline
    \noalign{\smallskip}
Input              & 872           & 0              \\
Dense              & 16            & 13698          \\
Dense              & 8             & 136            \\
Dense              & 4             & 36             \\
Dense              & 1             & 5              \\
    \noalign{\smallskip}
\hline
\end{tabular}
\end{table}
 
Once the configuration has been defined, the training process takes place. Due to the fact that convolutional layers are frozen, the evolution of models is due to the update of variable weights, which requires a larger number of iterations than for training the whole convolutional neural network (CNN) at once. Indeed, specific attention must be paid to the cross-validation strategy used to avoid over-fitting, which means that the same number of models as folders is required. Different numbers of folders were tested in the cross-validation process (from three to five), as well as different bases for features derived from different DL models, which required a significant number of repetitions for the training procedure before being able to propose a set of DTL models. In the present case, and because of the limited number of available samples, the best method for measuring the global quality of the whole data set was using four cross-validation folders, which were then adopted for the implementation. Therefore, the number of computing operations in the training step is expected to be large, and advanced computing capabilities are targeted to keep the effort bounded. As the operations are tensor-based computations, the use of existing frameworks an save a lot of time.

The adoption of the {\tt TensorFlow} framework \citep{abadi2016tensorflow} for the creation of DL models enables the use of accelerated hardware based on Nvidia general-purpose graphics processing unit (GPU) cards, which outperform the central processing unit (CPU) in terms of 
computation time by around a factor of 20 \citep{Mittal2019}.
As the base DL models were selected from those performed in~\cite{Passegger2020} and were able to identify the best features, the same framework was retained for the current implementation of DTL. 
Current features were extracted from the aforementioned models, and a complete new deep ANN was configured and trained over the new set of spectra with better stellar parameter estimation. In this particular case, because training involved only adjustment of the deep ANN weights, several thousand epochs were required to produce and estimate the adapted function.

In this application, we used GPU cards with 11\,GB of RAM and 4352 computing cores.
The training time for a model experiencing proper convergence depended on the training data size, but also on the architecture and number of epochs, and varied between 45 minutes and two hours.
For a more detailed description of the general design of a DL neural network, we refer to \citet{Passegger2020}. 

The same methodology used by~\cite{Passegger2020} for uncertainty estimation was considered here, where parameter estimations from each DTL model were collected and the probability density function was determined using KDE \citep{scott2015multivariate,terrell1992variable,wang2017robust}.
Based on such a probability {density} function, the maximum was retained as a confident estimation of the parameter. This was done for each star and stellar parameter separately. To provide the uncertainty for each star and
parameter, the $\pm$ 1~$\sigma$ thresholds of the predictions were calculated.

\section{Results and discussion}
\label{Results}

We introduced an algorithm-independent assessment of precision in the prediction of $T_{\rm eff}$ and [M/H]. This was carried out thanks to the testing data set, where stars not seen before by the model during its knowledge transference were used as the gold standard for its assessment for estimating stellar parameters.
First, we applied the DL method presented by \cite{Passegger2020} by selecting the best DL model that predicts $T_{\rm eff}$ from PHOENIX-ACES. In doing so, the hypothesis that the derived models use the most relevant feature set after the convolution step remains acceptable. This first step also provided a set of comparisons of stellar parameters derived from DL.


\begin{figure*}
  \centering
      \begin{minipage}{0.49\textwidth}
        \centering
        \includegraphics[width=0.99\textwidth]{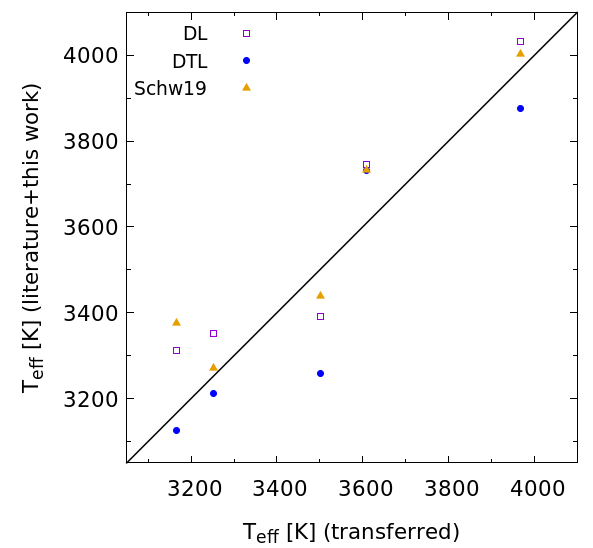}  
    \end{minipage}\hfill
    \begin{minipage}{0.49\textwidth}
        \centering
        \includegraphics[width=0.99\textwidth]{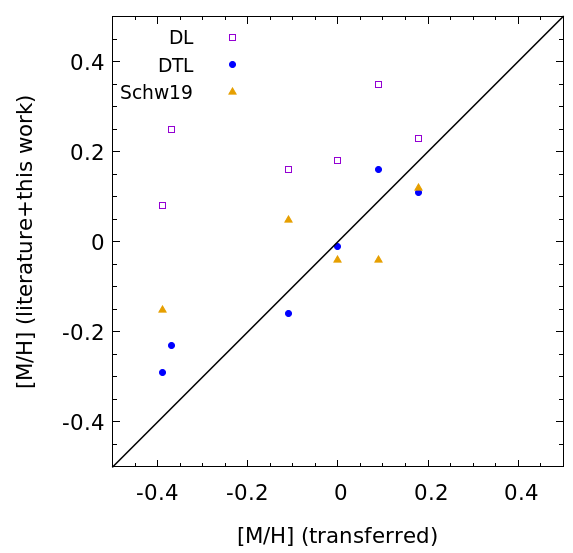}  
    \end{minipage}\hfill
    \caption{Comparison of $T_{\rm eff}$ ({\it left}) and [M/H] ({\it right}) from different techniques for stars not used in the DTL training process. Values are presented in Tables~\ref{tab:teff_uncert} and \ref{tab:zeta_uncert}.}
    \label{fig:dtl_ref_teff_zeta}
\end{figure*}

According to the steps indicated in Sect.~\ref{transferred.knowledge}, starting from the best DL model, the network-based transfer procedure was carried out within the samples from Tables~\ref{tab:interf_teff} {and~\ref{tab:lit_teff}.} In this procedure, 14 stars were used during the four-fold cross-validation approach, keeping aside the remaining five stars in {Table~\ref{tab:teff_uncert}}.
After creating the most suitable model, its quality was assessed by measuring the residual errors with the available information from interferometry.
The results for the five stars kept out {of} the training validation process of DTL were then used for independent quality assessment.
In the left panel of Fig.~\ref{fig:dtl_ref_teff_zeta}, we compare the interferometric values for these five stars to results from DTL, DL, and \citet[][Schw19]{Schweitzer2019}, as all these {studies} used PHOENIX-ACES models. 
For more than half of the stars, the accuracy of the DTL approach exceeds that of the DL and Schw19 approaches, while for only two stars (J15194--077 and J09144+526) the literature gives closer results to the interferometric values than DTL.

By using the same strategy as for $T_{\rm eff}$, [M/H] data from {Tables~\ref{tab:binaries_zeta} and} \ref{tab:lit_zeta} enabled us to use 18 stars for a four-fold cross-validation approach, keeping another 6 stars aside this time {(see Table~\ref{tab:zeta_uncert})}. 
The right panel of Fig.~\ref{fig:dtl_ref_teff_zeta} shows that the accuracy of the best deep transferred model for those six stars is significantly better than other existing estimations. For only one star (J22565+165), DL gives a closer parameter estimation to the transferred value than DTL. For the rest of the stars, DTL produces more accurate results than the other methods. 

Finally, we quantitatively assessed the accuracy improvements of our method with respect to previous implementations.
For metallicity, the median absolute differences (MADs) with respect to the transferred values are 0.07\,dex for DTL, 0.27\,dex for DL, and 0.13\,dex for Schw19. On the other hand, for $T_{\rm eff}$, Schw19 provides the smallest MAD with 60\,K. However, DTL outperforms DL with differences of 91\,K and 110\,K, respectively.

\subsection{Uncertainties}

Until this point, this paper proposes a technique to transfer the features identified as relevant according to the stellar parameter of interest and the selected flux window. This technique uses precise estimations of stellar properties from interferometry and binary observations to adapt the knowledge domain based on the previous DL feature identification.
{Our} technique {provides} good estimations for the stars used for quality control as it outperforms the reference estimations for the sample of selected stars in the majority of cases. 

When the same set of features is transferred through a different model to adapt the knowledge domain, a different subset of features (components of the feature vector) can be selected. 
This leads to different parameter estimations after a converged training process, which reflects the effects of the different components of the feature vector. Therefore, we defined uncertainties in the DTL process for estimating stellar parameters as $\pm 1 \sigma$ around the most frequently predicted value for each parameter. In this way, each star has its own uncertainty interval, {which does not have to be symmetric.}

\begin{figure*}
\centering
        \includegraphics[width=0.49\textwidth]{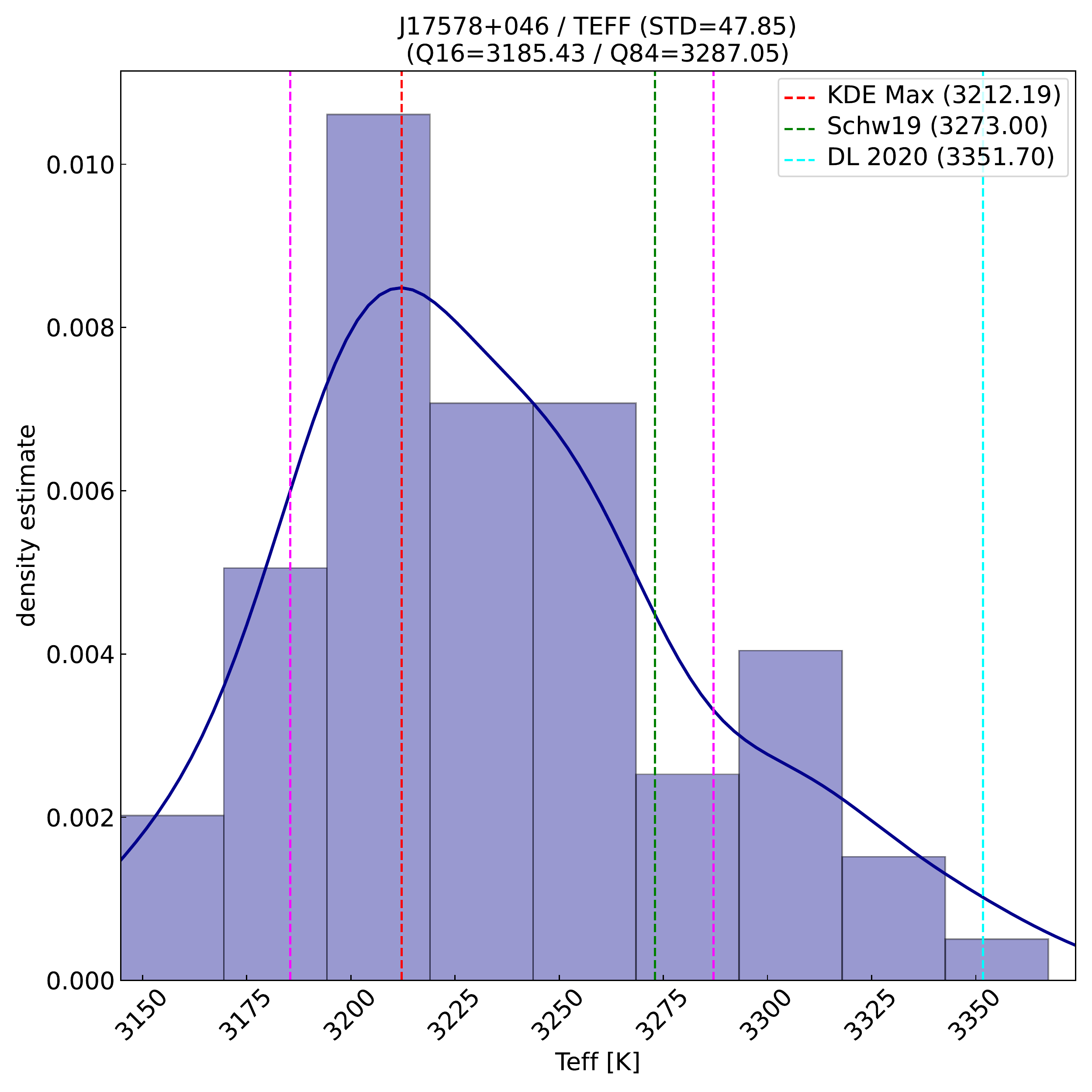}  
        \includegraphics[width=0.49\textwidth]{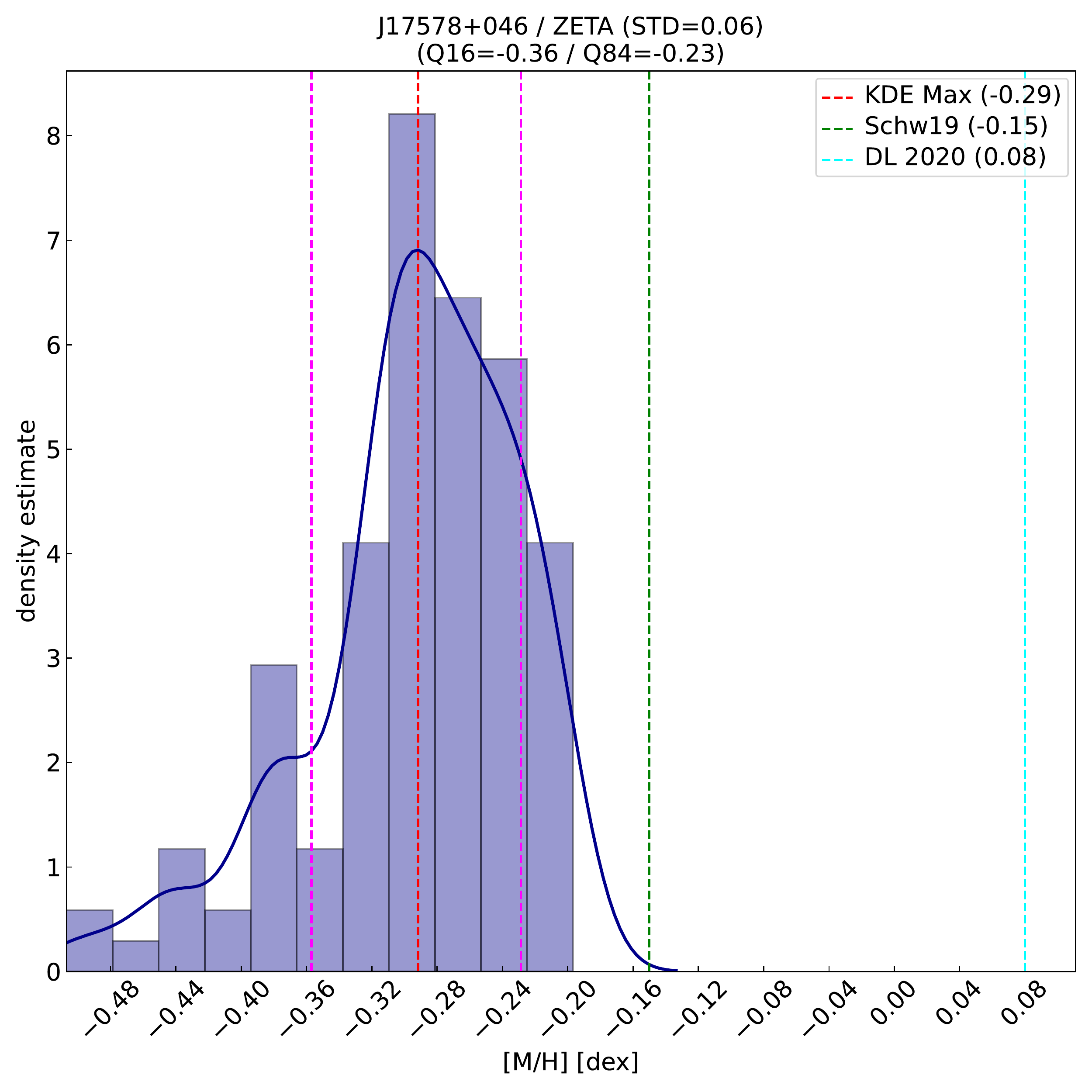}  
        \caption{Different DTL estimations of $T_{\rm eff}$  ({\it left}) and [M/H] ({\it right}) for a single star (J17578+046, Barnard’s Star, M3.5\,V).
        Quantiles $\pm 1 \sigma$ are indicated with dashed magenta lines. 
        Results from DL (cyan) and Schw19 (green) are also included for comparison.}
        \label{fig:dtl_teff_zeta_kde}
\end{figure*}

The DTL process can be repeated several times, making it possible to get different transferred models, each of them weighting the different extracted features differently. 
The proposal was to retain the models that are over a specific quality threshold and then to aggregate the {estimates} from them using an integrated KDE technique, as mentioned in Sect.~\ref{transferred.knowledge}. 
An example for $T_{\rm eff}$ and for [M/H] is shown in Fig.~\ref{fig:dtl_teff_zeta_kde}.
Indeed, the observed shapes allow us to discuss the number of DTL models considered to {produce} good-quality estimations. 
We selected 80 DTL models to be considered as {estimate} makers, which is a matter of design. However, for some stars, more evidence was needed in order to reduce the uncertainty level, as illustrated in the right half of the left panel in Fig.~\ref{fig:dtl_teff_zeta_kde}, where the tail increases the uncertainty value for {the 84\,\% quantile}. 
The estimated uncertainties for the selected quality stars are presented in Table~\ref{tab:teff_uncert} for $T_{\rm eff}$ and Table~\ref{tab:zeta_uncert} for [M/H] parameters.

\begin{table}[]
\caption{$T_{\rm eff}$ comparison for the stars not used in the training phase\tablefootmark{a}.}
\label{tab:teff_uncert}
\centering %
\begin{adjustbox}{width=0.49\textwidth}
\begin{tabular}{l cccc}
    \hline 
    \hline 
    \noalign{\smallskip}
\multicolumn{1}{c}{Karmn} & \multicolumn{1}{c}{$T_{\rm eff, interf}$} & \multicolumn{1}{c}{$T_{\rm eff, DL}$} & \multicolumn{1}{c}{$T_{\rm eff, DTL}$} & \multicolumn{1}{c}{$T_{\rm eff, Schw19}$}\\
 & [K] & [K] & [K] & [K] \\
\noalign{\smallskip}
\hline
\noalign{\smallskip}
J09144+526 & $3968 \pm 42$ & $4033 \pm 60$ & $3877 \pm 16$ & $4005 \pm 51$ \\
J15194--077 & $3501 \pm 58$ & $3391 \pm 47$ & $3259 \pm 24$ & $3441 \pm 51$ \\
J16581+257 & $3610 \pm 61$ & $3745 \pm 43$ & $3732 \pm 11$ & $3734 \pm 51$ \\
J17578+046 & $3252 \pm 23$ & $3352 \pm 55$ & $3212 \pm 48$ & $3273 \pm 51$ \\
J22532--142 & $3165 \pm 25$ & $3313 \pm 55$ & $3125 \pm 37$ & $3377 \pm 51$ \\
\noalign{\smallskip}
\hline
\end{tabular}
\end{adjustbox}
\tablefoot{
    \tablefoottext{a}{Observed (`interf'), DL-estimated (`DL'), DTL-predicted (`DTL'), and Schw19 values.}
}
\end{table}

\begin{table}[]
\caption{[M/H] comparison for the stars not used in the training phase.\tablefootmark{a}}
\label{tab:zeta_uncert}
\centering %
\begin{adjustbox}{width=0.49\textwidth}
\begin{tabular}{l cccc}
    \hline 
    \hline 
    \noalign{\smallskip}
\multicolumn{1}{c}{Karmn} & \multicolumn{1}{c}{[M/H]$_{\rm\,Bin,lit}$} & \multicolumn{1}{c}{[M/H]$_{\rm\,DL}$} & \multicolumn{1}{c}{[M/H]$_{\rm\,DTL}$} & 
\multicolumn{1}{c}{[M/H]$_{\rm\,Schw19}$} \\
 & [dex] & [dex] & [dex] & [dex]\\
\noalign{\smallskip}
\hline
\noalign{\smallskip}
J04153--076  & $-0.37 \pm 0.02$ & $+0.25 \pm 0.19$ & $-0.23 \pm 0.05$ & ... \\
J07361--031  & $-0.11 \pm 0.03$ & $+0.16 \pm 0.10$ & $-0.16 \pm 0.04$ & $+0.05 \pm 0.16$ \\
J11421+267  & $+0.00 \pm 0.11$ & $+0.18 \pm 0.09$ & $-0.01 \pm 0.03$ & $-0.04 \pm 0.16$ \\
J13005+056  & $+0.09 \pm 0.10$ & $+0.35 \pm 0.14$ & $+0.16 \pm 0.10$ & $-0.04 \pm 0.16$ \\
J17578+046  & $-0.39 \pm 0.12$ & $+0.08 \pm 0.14$ & $-0.29 \pm 0.06$ & $-0.15 \pm 0.16$ \\
J22565+165  & $+0.18 \pm 0.09$ & $+0.23 \pm 0.09$ & $+0.11 \pm 0.03$ & $+0.12 \pm 0.16$ \\
\noalign{\smallskip}
\hline
\end{tabular}
\end{adjustbox}
\tablefoot{
    \tablefoottext{a}{Observed (Bin,lit, i.e., [M/H] from primaries of FGK+M systems in the literature), DL-estimated (DL), and DTL-predicted (DTL) values. Uncertainties are their standard deviations.}
}
\end{table}

After testing the quality of the DTL models for estimating $T_{\rm eff}$ and [M/H], they were applied to the entire CARMENES data set. The outcome can be found in Table~\ref{tab:teff_zeta_carm}.
Stars with high rotational velocities {and therefore} with significantly higher uncertainties, were included even when the training data set did not contain this type of object.
In Table~\ref{tab:teff_zeta_carm}, there are several parameters with significantly small uncertainties, even smaller than those of the training sample (e.g., $\Delta T_{\rm eff}$ = 10\,K).
The uncertainties that we provide for the estimated stellar parameters refer to the internal error of the method, and therefore they do not take into account uncertainties from the interferometric and binary training samples or from the synthetic gap.

\begin{figure}[]
  \centering
  \includegraphics[width=0.49\textwidth]{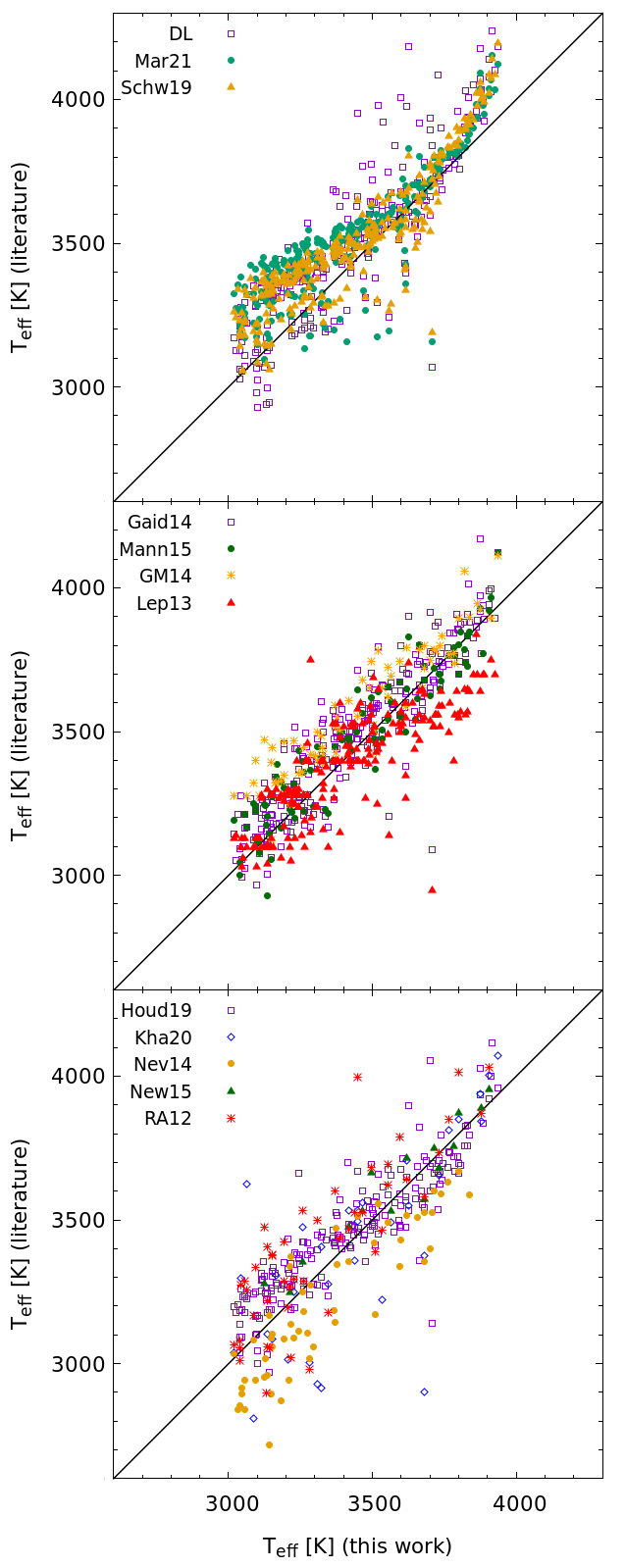}  
  \caption{Comparisons with estimations of $T_{\rm eff}$ from different techniques. The black line corresponds to the 1:1 relation.}
  \label{fig:dtl_cmp_teff}
\end{figure}

\begin{figure}[]
  \centering
  \includegraphics[width=0.49\textwidth]{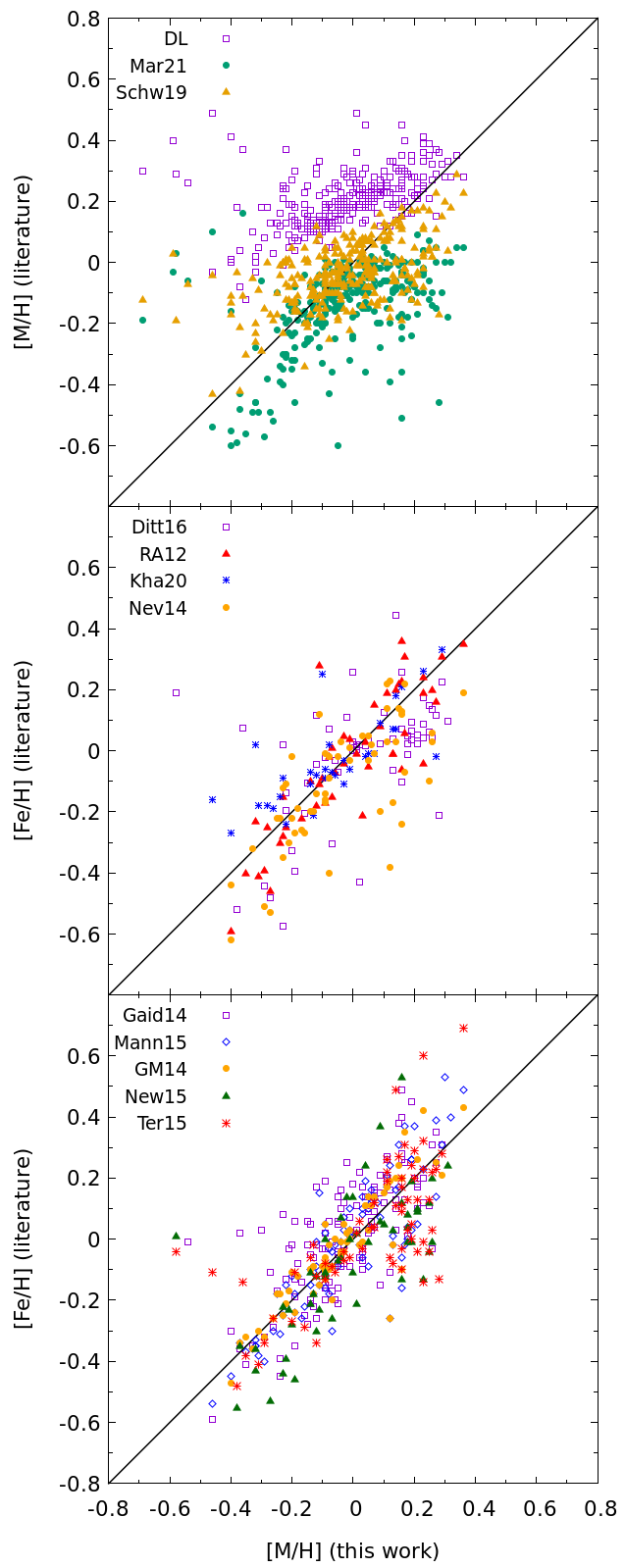}  
  \caption{Comparisons with estimations of [M/H] from different techniques. The black line corresponds to the 1:1 relation.}
  \label{fig:dtl_cmp_zeta}
\end{figure}

\subsection{Comparison with the literature}
We compare our DTL results for the full CARMENES sample for $T_{\rm eff}$ and [M/H] with other estimations from the literature. For improved readability, we divide the comparisons into three plots. We also present the Pearson correlation coefficient $r_{\rm P}$, which we use to assess the goodness of the correlation. 
A summary can be found in Table~\ref{tab:pearson_lit}.

\begin{table}[]
\caption{Summary of mean difference ($\overline{\Delta}$), standard deviation (std. dev.), and Pearson correlation coefficient ($r_{\rm P}$) for the comparison between literature values and DTL results.}
\label{tab:pearson_lit}
\centering %
\begin{tabular}{l ccc}
    \hline 
    \hline 
    \noalign{\smallskip}
    Reference & $\overline{\Delta}$ & std. dev. &  $r_{\rm P}$ \\
            &   $T_{\rm eff}$ / [M/H] & $T_{\rm eff}$ / [M/H] & $T_{\rm eff}$ / [M/H]\\
            & [K] / [dex] & [K] / [dex]
            & [K] / [dex] \\
\noalign{\smallskip}
\hline
\noalign{\smallskip}
DL  & +83 / +0.23 & 133 / 0.16 & 0.87 / 0.45 \\
Mar21  & +97 / -0.10 & 119 / 0.17 & 0.88/ 0.47 \\
Schw19  & +85 / --0.01 & 110 / 0.15 & 0.90 / 0.54 \\
GM14  & +121 / +0.01 & 96 / 0.08 & 0.92 / 0.89 \\
Mann15  & +33 / +0.02 & 92 / 0.11 & 0.94 / 0.87 \\
Gaid14  & +54 / +0.04 & 137 / 0.14 & 0.86 / 0.72 \\
Lep13  & --23 / ... & 138 / ... & 0.81 / ... \\
Houd19  & +40 / ... & 113 / ... & 0.89 / ... \\
Kha20  & --59 / --0.08 & 261 / 0.12 & 0.73 / 0.81 \\
Nev14  & --132 / --0.04 & 119 / 0.15 & 0.89 / 0.71 \\
New15  & +48 / --0.06 & 74 / 0.17 & 0.95 / 0.69 \\
RA12  & +67 / +0.01 & 160 / 0.10 & 0.83 / 0.86 \\
Ditt16  & ... / --0.05 & ... / 0.20 & ... / 0.50 \\
Ter15  & ... / --0.03 & ... / 0.17 & ... / 0.69 \\
\noalign{\smallskip}
\hline
\end{tabular}
\end{table}

Figure~\ref{fig:dtl_cmp_teff} presents the comparison with literature values for $T_{\rm eff}$. The top panel shows results from DL, \citet[][Mar21]{marfil2021carmenes}, and Schw19. All three studies used CARMENES data and show a similar pattern, with hotter temperature for stars below 3500\,K and above 3800\,K compared to DTL. For DL, the dispersion is larger than for the other two methods, also showing a slightly smaller $r_{\rm P}$ of 0.87 compared to 0.88 for Mar21 and 0.90 for Schw19. The main factor leading to this effect in DTL is the limited number of samples with temperatures above 4000\,K in the training and validation sets. Another possible explanation could be a change in opacity at around 3300~K, but a detailed analysis of the synthetic model structures would be necessary to come to any robust conclusions.

In the middle panel, all literature references determined $T_{\rm eff}$ by fitting BT-Settl synthetic models \citep{Allard2011} to VIS spectra. \citet[][GM14]{GaidosMann2014} additionally used spectral curvature indices in the $K$ band if there were no VIS spectra available. {Perhaps} this difference explains a general trend towards hotter temperatures compared to DTL (on average +121\,K, with $r_{\rm P}$=0.92). \citet[][Gaid14]{Gaidos2014} and \citet[][Mann15]{Mann2015} achieve similar results to with DTL, $r_{\rm P}$=0.86 and 0.94, respectively. The correlation of \citet[][Lep13]{Lepine2013} is weaker with $r_{\rm P}$=0.81, providing some cooler values at the hot end of the $T_{\rm eff}$ scale. 

For the comparison in the bottom panel, literature values were determined from the H$_{2}$O--K2 index as defined by \citet{RojasAyala2012}, empirical relations with EWs, and, in the case of \citet[][Houd19]{Houdebine2019}, photometric relations. Results from Houd19 and \citet[][New15]{Newton2015} show a good correlation with DTL ($r_{\rm P}$ = 0.89 and 0.95, respectively), while \citet[][RA12]{RojasAyala2012} is slightly on the hotter side of the 1:1 relation with $r_{\rm P}$ = 0.83. \citet[][Kha20]{Khata2020} exhibits a large spread ($r_{\rm P}$ = 0.73), and values from \citet[][Nev14]{Neves2014} are on average 130\,K cooler than DTL, but still well correlated with $r_{\rm P}$ = 0.89. Nev14 measured pEWs of most lines and features in the optical and then used reference photometric scales of $T_{\rm eff}$ and [Fe/H] from \cite{Casagrande2008} and \cite{Neves2012}, respectively. Both scales are based on [Fe/H] determinations from FGK+M binaries. The offset that we see between the findings of these latter authors and our $T_{\rm eff}$ values also exists compared to other literature works, which indicates an intrinsic underestimation of temperatures by Nev14. Kha20 determined EWs of Mg (1.57\,$\mu$m, 1.71\,$\mu$m), Al (1.67\,$\mu$m), and the H$_{2}$O index in the $H$ band, and performed linear regression on 12 M-dwarf calibrator stars with interferometrically measured $T_{\rm eff}$ to derive a temperature relation. The standard deviation of the residuals of the calibrators amounted to 102\,K. However, the spread with respect to our results and other literature studies is much larger, with a standard deviation of 261 \,K. An indication of a higher deviation in the comparison with the literature can already be seen in Fig.~5 of Kha20, where these authors compare their results with $T_{\rm eff}$ from Mann15 and RA12, and measure standard deviations of 164\,K and 158\,K, respectively. At this point, we cannot give a clear explanation for this behavior.
Overall, the correlation between DTL and the literature is quite good, except for Mar21, Schw19, and Kha20, where we are not sure about the source of the differences seen.

Figure~\ref{fig:dtl_cmp_zeta} presents the same comparisons for metallicity. As explained by \cite{Passegger2020}, our derived [M/H] values directly translate into identical [Fe/H] values. Therefore, we can compare [M/H] values with [Fe/H] from the literature. Furthermore, [Fe/H] is often used as a proxy for [M/H] in the literature.

The top panel again shows the results from DL, Mar21, and Schw19. There is an offset towards more metal-rich values for DL (on average 0.23\,dex, $r_{\rm P}$ = 0.45), and towards more metal-poor values for Mar21 (on average $-0.11$\,dex, $r_{\rm P}$ = 0.46). A possible explanation for these offsets might be that Mar21 reported [Fe/H] corrected for $\alpha$-enhancement. A trend of more metal-rich values for DL was already shown by \cite{Passegger2020,Passegger2022}. The results from Schw19 are consistent overall with DTL, with $r_{\rm P}$ = 0.54, although they exhibit a certain spread.

The middle panel of Fig.~\ref{fig:dtl_cmp_zeta} compares similar determination methods from the literature. RA12 and Kha20 derived [Fe/H] using the H$_{2}$O--K2 index and equivalent widths (EWs) of Na~{\sc i} and Ca~{\sc i} in the NIR. Nev14 also incorporated a relation with EWs, but only \citet[][Ditt16]{Dittmann2016} determined [Fe/H] from a color--magnitude--metallicity relation. This might explain the large spread of the latter values, {resulting in} $r_{\rm P}$ = 0.50. Results from Nev14 are overall more metal-poor than those provided by DTL and show a large spread as well, but a better correlation than Ditt16, with $r_{\rm P}$ = 0.71. However, values from RA12 and Kha20 correspond even better with DTL, showing $r_{\rm P}$ = 0.86 and 0.81, respectively.

The literature values shown in the bottom panel of Fig. 10 were determined using empirical relations between atomic line strength, Na~{\sc i} and Ca~{\sc i} EWs, the H$_{2}$O--K2 index, and metallicity calibrated with FGK+M binaries based on \cite{Mann2013a,Mann2013b,Mann2014} relationships. The values provided by GM14, Mann15, and Gaid14 are highly correlated, with those of Gaid14 showing the least spread ($r_{\rm P}$ = 0.89, 0.87, and 0.72, respectively). {Values from} New15 are generally slightly more metal-poor, with a smaller correlation coefficient of $r_{\rm P}$ = 0.69. For higher metallicities, \citet[][Ter15]{Terrien2015} {derived} a larger spread and some outliers at both ends. The correlation coefficient is the same as for New15.
Similar to $T_{\rm eff}$, DTL values for metallicity correspond well with most of the literature, and an improvement with respect to DL can be appreciated, which is very promising.


\section{Summary and conclusions}
\label{Summary}

We present a DTL neural network technique that improves the estimation of the stellar parameters $T_{\rm eff}$ and [M/H] for M dwarfs from high-S/N, high-resolution optical spectroscopy obtained with CARMENES. The initial DL model was trained with PHOENIX-ACES synthetic spectra, which confer the advantage that they allow a sufficient number of spectra to be generated with known stellar parameters. Based on the DL convolutional features, different DTL models were trained and tested. To use a more robust procedure, a cross-validation scheme was adopted. Using the proposed technique could help to bridge the synthetic gap affecting stellar parameter estimation based on synthetic libraries. However, a larger stellar sample covering a wider spectral range is needed to verify this.

Before applying the created models to a large data set, we defined an independent quality assessment procedure based on specific stars for which high-quality stellar parameter estimations are available.
This assessment shows that the DTL technique has good prediction capabilities. 
In addition, we incorporated an uncertainty estimation procedure based on considering the diversity of {estimates} from different transferred models, as well as an aggregation procedure. Such an estimation is flux-window dependent, but also star dependent, because the trained DTL models below the convergence threshold depend on them.

Another relevant aspect to be considered for parameter estimation concerns the selected flux windows, as they have their own influence {on} the vector of features, and, in the end, on the estimated parameters. A possible continuation of the line of research in this field is applying reinforced learning techniques based on the behavior of the selected flux windows.

Finally, and importantly, a limitation of the proposed method is the parameter range of the observed transferred knowledge. This means that the {parameters of} stars with large $\varv\sin{i}$ values {cannot} be estimated rigorously with this technique, as no interferometric or FGK-star companion values are available with such large values of rotational velocity. In the same way, the transferred knowledge works only for $T_{\rm eff}$ higher than 3100\,K, as no cooler stars were part of the training set yet. Therefore, our current analysis is limited to spectral types between M0\,V and M6\,V.

In summary, we propose an innovative technique that can increase the value of its predictions as new high-quality stellar parameters ---namely $T_{\rm eff}$ from interferometry and [M/H] from FGK+M systems--- become available in the near future. The current data sample used is close to the operational limits of the technique, while in some cases the data set was {complemented} with stellar parameters estimated from the literature. Therefore, improvements are expected to be possible when more stars with highly reliable stellar parameters and high-resolution spectra become available. In the meantime, the lack of a sufficiently large number of samples is a limitation for the technique.


\begin{acknowledgements}
We thank an anonymous referee for helpful comments that improved the quality of this paper.
CARMENES is an instrument at the Centro Astron\'omico Hispano en Andaluc\'ia (CAHA) at Calar Alto (Almer\'{\i}a, Spain), operated jointly by the Junta de Andaluc\'ia and the Instituto de Astrof\'isica de Andaluc\'ia (CSIC).
CARMENES was funded by the German Max-Planck-Gesellschaft (MPG), the Spanish Consejo Superior de Investigaciones Cient\'ificas (CSIC), European Regional Development Fund (ERDF) through projects FICTS-2011-02, ICTS-2017-07-CAHA-4, and CAHA16-CE-3978, 
and the members of the CARMENES Consortium (Max-Planck-Institut f\"ur Astronomie, Instituto de Astrof\'isica de Andaluc\'ia, Landessternwarte K\"onigstuhl, Institut de Ci\`encies de l'Espai, Insitut f\"ur Astrophysik G\"ottingen, Universidad Complutense de Madrid, Th\"uringer Landessternwarte Tautenburg, Instituto de Astrof\'isica de Canarias, Hamburger Sternwarte, Centro de Astrobiolog\'ia and Centro Astron\'omico Hispano-Alem\'an), with additional contributions by the Spanish Ministry of Economy, the German Science Foundation through the Major Research Instrumentation Programme and DFG Research Unit FOR2544 ``Blue Planets around Red Stars'', the Klaus Tschira Stiftung, the states of Baden-W\"urttemberg and Niedersachsen, and by the Junta de Andaluc\'ia.

We also acknowledge the financial support from MCIN/AEI/10.13039/501100011033/ and the ERDF through projects 
PID2021-125627OB-C31, 
PID2020-112949GB-I00, 
and PID2019-109522GB-C51/2/3/4, 
and fellowship FPU15/01476; 
the Funda\c{c}\~{a}o para a Ci\^{e}ncia e a Tecnologia through and ERDF through grants UID/FIS/04434/2019, UIDB/04434/2020, and UIDP/04434/2020, PTDC/FIS-AST/28953/2017; 
grant RTI2018-094614-B-I00 (SMASHING) into the ``Programa Estatal de I+D+i Orientada a los Retos de la Sociedad'' funded by MCIN/AEI/10.13039/501100011033, and COMPETE2020 - Programa Operacional Competitividade e Internacionaliza\c{c}\~{a}o POCI-01-0145-FEDER-028953;
the University of La Laguna and the EU Next Generation funds through the Margarita Salas Fellowship from the Spanish Ministerio de Universidades UNI/551/2021-May 26;
and NASA through grant NNX17AG24G.

\end{acknowledgements}
\bibliographystyle{aa}
\bibliography{CARM_DTL}

\begin{appendix}
\section{DTL estimations for CARMENES survey stars}
\captionof{table}{$T_{\rm eff}$ and [M/H] values for CARMENES stars estimated by the DTL method\tablefootmark{a}.}
\tablefirsthead{\toprule \midrule Karmn & $T_{\rm eff}$ [K] & [M/H] [dex] \\ \midrule}
\tablehead{
\multicolumn{3}{c}
{\tablename\ \thetable\ -- \textit{Continued from previous page} } \\
\toprule
\midrule
Karmn & $T_{\rm eff}$ [K] & [M/H] [dex] \\ \midrule}
\tabletail{
\midrule \multicolumn{3}{r} {\vspace{3pt} \textit{Continued on next page}} \\ \midrule 
}
\tablelasttail{
\bottomrule {}
}
\xentrystretch{0.09}
\setlength{\extrarowheight}{3pt}
\begin{center}
\begin{xtabular}{lll}
\label{tab:teff_zeta_carm}
JJ00051+457 & $3721_{-8}^{+8}$ & $-0.02_{-0.02}^{+0.04}$ \\
J00067--075$^{1,2}$ & $3140_{-38}^{+249}$ & $-0.19_{-0.03}^{+0.28}$ \\
J00162+198E & $3149_{-31}^{+46}$ & $-0.02_{-0.03}^{+0.06}$ \\
J00183+440$^{1,2}$ & $3563_{-9}^{+1}$ & $-0.26_{-...}^{+...}$ \\
J00184+440 & $3219_{-41}^{+71}$ & $-0.07_{-0.10}^{+0.06}$ \\
J00286--066 & $3246_{-22}^{+40}$ & $-0.06_{-0.04}^{+0.02}$ \\
J00389+306 & $3503_{-13}^{+10}$ & $-0.09_{-0.03}^{+0.02}$ \\
J00570+450 & $3341_{-19}^{+30}$ & $-0.09_{-0.04}^{+0.01}$ \\
J01013+613 & $3489_{-16}^{+20}$ & $-0.18_{-0.02}^{+0.03}$ \\
J01025+716 & $3449_{-18}^{+33}$ & $-0.03_{-0.05}^{+0.03}$ \\
J01026+623 & $3743_{-8}^{+9}$ & $+0.10_{-0.01}^{+0.06}$ \\
J01048--181 & $3101_{-23}^{+160}$ & $+0.17_{-0.08}^{+0.04}$ \\
J01125--169 & $3147_{-49}^{+224}$ & $+0.12_{-0.15}^{+0.08}$ \\
J01339--176 & $3279_{-23}^{+118}$ & $+0.04_{-0.06}^{+0.06}$ \\
J01433+043 & $3514_{-14}^{+8}$ & $-0.09_{-0.03}^{+0.01}$ \\
J01518+644 & $3570_{-8}^{+16}$ & $+0.06_{-0.01}^{+0.05}$ \\
J02002+130 & $3109_{-52}^{+244}$ & $+0.16_{-0.13}^{+0.06}$ \\
J02015+637 & $3457_{-8}^{+14}$ & $+0.06_{-0.02}^{+0.03}$ \\
J02070+496 & $3423_{-31}^{+41}$ & $-0.12_{-0.03}^{+0.04}$ \\
J02088+494$^{3}$ & $3476_{-88}^{+200}$ & $-0.40_{-0.11}^{+0.12}$ \\
J02123+035 & $3520_{-10}^{+19}$ & $-0.33_{-0.01}^{+0.01}$ \\
J02222+478 & $3819_{-5}^{+8}$ & $+0.05_{-0.02}^{+0.05}$ \\
J02336+249 & $3230_{-38}^{+177}$ & $+0.18_{-0.09}^{+0.04}$ \\
J02358+202 & $3650_{-9}^{+11}$ & $+0.08_{-0.01}^{+0.06}$ \\
J02362+068$^{2}$ & $3147_{-22}^{+46}$ & $-0.20_{-...}^{+...}$ \\
J02442+255 & $3321_{-14}^{+20}$ & $-0.14_{-0.04}^{+0.01}$ \\
J02519+224$^{3}$ & $3410_{-116}^{+205}$ & $-0.59_{-0.22}^{+0.12}$ \\
J02565+554W & $3823_{-9}^{+8}$ & $+0.16_{-0.02}^{+0.07}$ \\
J03133+047 & $3048_{-12}^{+109}$ & $+0.25_{-0.02}^{+0.04}$ \\
J03181+382 & $3799_{-7}^{+9}$ & $+0.30_{-0.06}^{+0.02}$ \\
J03213+799 & $3513_{-16}^{+12}$ & $-0.10_{-0.03}^{+0.02}$ \\
J03217--066 & $3615_{-16}^{+14}$ & $-0.03_{-0.02}^{+0.05}$ \\
J03463+262 & $3890_{-14}^{+5}$ & $+0.07_{-0.01}^{+0.06}$ \\
J03473--019 & $3612_{-24}^{+93}$ & $-0.14_{-0.07}^{+0.05}$ \\
J03531+625 & $3447_{-11}^{+15}$ & $-0.18_{-0.02}^{+0.01}$ \\
J04153--076 & $3282_{-57}^{+132}$ & $-0.23_{-0.05}^{+0.06}$ \\
J04225+105 & $3277_{-18}^{+29}$ & $+0.14_{-0.03}^{+0.04}$ \\
J04290+219 & $3936_{-20}^{+11}$ & $+0.27_{-0.07}^{+0.03}$ \\
J04311+589 & $3042_{-35}^{+40}$ & $+0.16_{-0.06}^{+0.02}$ \\
J04376+528 & $3876_{-7}^{+6}$ & $-0.13_{-0.04}^{+0.01}$ \\
J04376--110 & $3595_{-12}^{+12}$ & $-0.04_{-0.02}^{+0.04}$ \\
J04429+189$^{1,2}$ & $3679_{-17}^{+2}$ & $+0.07_{-0.01}^{+0.05}$ \\
J04429+214 & $3328_{-25}^{+23}$ & $+0.07_{-0.02}^{+0.05}$ \\
J04520+064 & $3193_{-28}^{+33}$ & $+0.16_{-0.06}^{+0.02}$ \\
J04538--177 & $3534_{-15}^{+7}$ & $-0.19_{-0.03}^{+0.01}$ \\
J04588+498 & $3861_{-14}^{+7}$ & $+0.06_{-0.02}^{+0.06}$ \\
J05019+011 & $3330_{-33}^{+209}$ & $+0.12_{-0.09}^{+0.04}$ \\
J05019--069 & $3141_{-33}^{+156}$ & $+0.16_{-0.14}^{+0.04}$ \\
J05033--173 & $3286_{-25}^{+26}$ & $-0.18_{-0.03}^{+0.01}$ \\
J05062+046$^{3}$ & $3263_{-87}^{+276}$ & $-0.46_{-0.19}^{+0.10}$ \\
J05127+196 & $3488_{-9}^{+19}$ & $-0.10_{-0.04}^{+0.02}$ \\
J05280+096 & $3216_{-14}^{+35}$ & $-0.20_{-0.02}^{+0.04}$ \\
J05314--036$^{1,2}$ & $3801_{-2}^{+9}$ & $+0.36_{-0.06}^{+0.01}$ \\
J05337+019 & $3614_{-28}^{+127}$ & $-0.30_{-0.05}^{+0.07}$ \\
J05348+138 & $3303_{-24}^{+20}$ & $-0.02_{-0.02}^{+0.02}$ \\
J05360--076 & $3076_{-23}^{+42}$ & $+0.12_{-0.04}^{+0.03}$ \\
J05365+113 & $3911_{-12}^{+7}$ & $-0.09_{-0.03}^{+0.02}$ \\
J05366+112 & $3283_{-37}^{+149}$ & $+0.08_{-0.10}^{+0.06}$ \\
J05415+534$^{2}$ & $3765_{-6}^{+18}$ & $+0.04_{-0.01}^{+0.06}$ \\
J05421+124 & $3085_{-17}^{+54}$ & $-0.22_{-0.01}^{+0.03}$ \\
J06000+027 & $3345_{-67}^{+179}$ & $+0.09_{-0.12}^{+0.05}$ \\
J06011+595 & $3164_{-12}^{+38}$ & $+0.05_{-0.04}^{+0.02}$ \\
J06024+498 & $3040_{-21}^{+186}$ & $+0.13_{-0.11}^{+0.07}$ \\
J06103+821 & $3507_{-11}^{+18}$ & $-0.13_{-0.03}^{+0.01}$ \\
J06105--218 & $3762_{-5}^{+12}$ & $+0.05_{-0.01}^{+0.06}$ \\
J06246+234 & $3182_{-51}^{+53}$ & $-0.19_{-0.06}^{+0.07}$ \\
J06371+175 & $3662_{-17}^{+11}$ & $-0.37_{-0.02}^{+0.02}$ \\
J06396--210 & $3157_{-19}^{+69}$ & $+0.03_{-0.05}^{+0.03}$ \\
J06421+035 & $3365_{-12}^{+21}$ & $-0.05_{-0.04}^{+0.01}$ \\
J06548+332 & $3308_{-15}^{+16}$ & $-0.06_{-0.04}^{+0.01}$ \\
J06574+740$^{3}$ & $3518_{-107}^{+184}$ & $-0.54_{-0.23}^{+0.12}$ \\
J06594+193 & $3100_{-38}^{+215}$ & $+0.04_{-0.06}^{+0.13}$ \\
J07033+346 & $3255_{-40}^{+156}$ & $+0.19_{-0.09}^{+0.04}$ \\
J07044+682 & $3378_{-11}^{+15}$ & $+0.00_{-0.02}^{+0.03}$ \\
J07274+052 & $3226_{-23}^{+36}$ & $-0.09_{-0.03}^{+0.02}$ \\
J07287--032 & $3369_{-16}^{+18}$ & $-0.04_{-0.02}^{+0.03}$ \\
J07319+362N & $3322_{-39}^{+126}$ & $+0.14_{-0.11}^{+0.06}$ \\
J07353+548 & $3521_{-20}^{+12}$ & $-0.19_{-0.02}^{+0.01}$ \\
J07361--031 & $3802_{-9}^{+17}$ & $-0.16_{-0.03}^{+0.03}$ \\
J07386--212 & $3262_{-17}^{+43}$ & $-0.25_{-0.03}^{+0.04}$ \\
J07393+021 & $3886_{-10}^{+5}$ & $+0.06_{-0.01}^{+0.06}$ \\
J07472+503$^{3}$ & $3263_{-28}^{+294}$ & $-0.04_{-0.08}^{+0.02}$ \\
J07558+833$^{1,2,3}$ & $3219_{-13}^{+335}$ & $-0.00_{-0.06}^{+0.01}$ \\
J07582+413 & $3115_{-14}^{+16}$ & $+0.04_{-0.04}^{+0.02}$ \\
J08119+087 & $3215_{-41}^{+91}$ & $-0.27_{-0.11}^{+0.06}$ \\
J08126--215 & $3034_{-23}^{+26}$ & $+0.16_{-0.04}^{+0.02}$ \\
J08161+013 & $3508_{-12}^{+20}$ & $-0.09_{-0.05}^{+0.01}$ \\
J08293+039 & $3614_{-11}^{+8}$ & $+0.05_{-0.01}^{+0.05}$ \\
J08315+730 & $3146_{-19}^{+52}$ & $-0.15_{-0.02}^{+0.02}$ \\
J08358+680 & $3435_{-19}^{+19}$ & $-0.11_{-0.04}^{+0.01}$ \\
J08402+314 & $3194_{-13}^{+38}$ & $-0.09_{-0.02}^{+0.03}$ \\
J08409--234 & $3273_{-28}^{+36}$ & $+0.11_{-0.04}^{+0.04}$ \\
J08413+594 & $3133_{-53}^{+238}$ & $+0.01_{-0.06}^{+0.16}$ \\
J08526+283$^{2}$ & $3063_{-26}^{+32}$ & $+0.29_{-0.04}^{+0.01}$ \\
J09005+465 & $3136_{-26}^{+128}$ & $+0.23_{-0.06}^{+0.04}$ \\
J09028+680 & $3205_{-28}^{+33}$ & $-0.16_{-0.02}^{+0.02}$ \\
J09133+688 & $3646_{-16}^{+13}$ & $+0.00_{-0.02}^{+0.05}$ \\
J09140+196 & $3486_{-24}^{+37}$ & $+0.07_{-0.02}^{+0.04}$ \\
J09143+526$^{1,2}$ & $3907_{-8}^{+3}$ & $-0.12_{-0.02}^{+0.00}$ \\
J09144+526$^{2}$ & $3877_{-8}^{+4}$ & $-0.07_{-0.04}^{+0.01}$ \\
J09161+018$^{3}$ & $3386_{-38}^{+241}$ & $-0.13_{-0.08}^{+0.05}$ \\
J09163--186 & $3671_{-17}^{+7}$ & $-0.05_{-0.02}^{+0.04}$ \\
J09307+003 & $3243_{-18}^{+29}$ & $-0.08_{-0.03}^{+0.01}$ \\
J09360--216 & $3377_{-14}^{+14}$ & $-0.21_{-0.03}^{+0.02}$ \\
J09411+132 & $3617_{-8}^{+10}$ & $-0.04_{-0.02}^{+0.04}$ \\
J09423+559 & $3230_{-11}^{+52}$ & $+0.00_{-0.02}^{+0.06}$ \\
J09425+700 & $3593_{-27}^{+18}$ & $+0.10_{-0.02}^{+0.07}$ \\
J09428+700 & $3424_{-23}^{+51}$ & $+0.13_{-0.07}^{+0.03}$ \\
J09439+269 & $3232_{-21}^{+30}$ & $-0.04_{-0.05}^{+0.02}$ \\
J09447--182 & $3127_{-18}^{+38}$ & $+0.03_{-0.04}^{+0.03}$ \\
J09468+760 & $3600_{-12}^{+18}$ & $-0.20_{-0.01}^{+0.02}$ \\
J09511--123 & $3699_{-14}^{+14}$ & $-0.19_{-0.03}^{+0.02}$ \\
J09561+627 & $3865_{-7}^{+5}$ & $+0.04_{-0.01}^{+0.05}$ \\
J10023+480 & $3729_{-12}^{+23}$ & $-0.05_{-0.05}^{+0.01}$ \\
J10122--037 & $3700_{-5}^{+10}$ & $+0.06_{-0.01}^{+0.06}$ \\
J10125+570 & $3238_{-19}^{+39}$ & $-0.05_{-0.03}^{+0.02}$ \\
J10167--119 & $3499_{-11}^{+28}$ & $+0.07_{-0.02}^{+0.04}$ \\
J10196+198 & $3510_{-23}^{+109}$ & $-0.11_{-0.08}^{+0.06}$ \\
J10251--102 & $3736_{-8}^{+4}$ & $+0.02_{-0.01}^{+0.05}$ \\
J10289+008 & $3598_{-17}^{+7}$ & $-0.13_{-0.03}^{+0.01}$ \\
J10350--094 & $3273_{-14}^{+34}$ & $-0.02_{-0.04}^{+0.02}$ \\
J10396--069 & $3543_{-19}^{+10}$ & $+0.07_{-0.02}^{+0.04}$ \\
J10416+376 & $3157_{-36}^{+44}$ & $-0.12_{-0.03}^{+0.08}$ \\
J10504+331 & $3245_{-23}^{+50}$ & $+0.16_{-0.06}^{+0.03}$ \\
J10508+068$^{1,2}$ & $3093_{-6}^{+21}$ & $+0.14_{-0.06}^{+0.01}$ \\
J11000+228 & $3438_{-13}^{+9}$ & $-0.10_{-0.04}^{+0.01}$ \\
J11026+219 & $3793_{-8}^{+13}$ & $-0.09_{-0.02}^{+0.04}$ \\
J11033+359$^{1,2}$ & $3465_{-2}^{+14}$ & $-0.31_{-...}^{+...}$ \\
J11054+435$^{1,2}$ & $3497_{-3}^{+4}$ & $-0.35_{-...}^{+...}$ \\
J11110+304W & $3712_{-7}^{+9}$ & $+0.13_{-0.01}^{+0.05}$ \\
J11126+189 & $3727_{-8}^{+6}$ & $+0.03_{-0.01}^{+0.05}$ \\
J11201--104 & $3702_{-16}^{+43}$ & $-0.15_{-0.06}^{+0.03}$ \\
J11289+101 & $3231_{-22}^{+26}$ & $-0.23_{-0.02}^{+0.04}$ \\
J11302+076 & $3426_{-16}^{+24}$ & $+0.05_{-0.02}^{+0.04}$ \\
J11306--080 & $3267_{-21}^{+30}$ & $-0.05_{-0.02}^{+0.04}$ \\
J11417+427 & $3192_{-16}^{+57}$ & $-0.03_{-0.04}^{+0.05}$ \\
J11421+267$^{1}$ & $3416_{-1}^{+4}$ & $-0.01_{-0.03}^{+0.03}$ \\
J11467--140 & $3575_{-13}^{+23}$ & $+0.27_{-0.07}^{+0.01}$ \\
J11476+786 & $3207_{-25}^{+36}$ & $-0.28_{-0.03}^{+0.03}$ \\
J11477+008 & $3019_{-30}^{+69}$ & $+0.13_{-0.12}^{+0.04}$ \\
J11509+483 & $3100_{-33}^{+89}$ & $+0.19_{-0.09}^{+0.04}$ \\
J11511+352 & $3664_{-12}^{+7}$ & $-0.14_{-0.03}^{+0.01}$ \\
J12054+695 & $3149_{-20}^{+67}$ & $+0.10_{-0.03}^{+0.08}$ \\
J12100--150 & $3170_{-35}^{+48}$ & $+0.19_{-0.07}^{+0.03}$ \\
J12111--199 & $3310_{-16}^{+25}$ & $-0.07_{-0.05}^{+0.01}$ \\
J12123+544S & $3852_{-9}^{+9}$ & $-0.09_{-0.03}^{+0.01}$ \\
J12156+526$^{3}$ & $3616_{-199}^{+121}$ & $-0.69_{-0.30}^{+0.13}$ \\
J12230+640 & $3598_{-36}^{+12}$ & $-0.06_{-0.02}^{+0.04}$ \\
J12248--182 & $3372_{-26}^{+12}$ & $-0.40_{-0.04}^{+0.02}$ \\
J12312+086 & $3829_{-5}^{+9}$ & $-0.07_{-0.03}^{+0.01}$ \\
J12350+098 & $3578_{-15}^{+15}$ & $-0.11_{-0.04}^{+0.02}$ \\
J12388+116 & $3420_{-26}^{+27}$ & $+0.19_{-0.06}^{+0.03}$ \\
J12428+418 & $3326_{-37}^{+130}$ & $+0.20_{-0.10}^{+0.04}$ \\
J12479+097 & $3211_{-22}^{+37}$ & $-0.04_{-0.03}^{+0.02}$ \\
J13005+056$^{1,3}$ & $3708_{-314}^{-3}$ & $+0.16_{-0.11}^{+0.08}$ \\
J13196+333 & $3734_{-11}^{+12}$ & $+0.32_{-0.07}^{+0.01}$ \\
J13209+342 & $3701_{-17}^{+13}$ & $-0.06_{-0.03}^{+0.02}$ \\
J13229+244 & $3206_{-38}^{+37}$ & $-0.22_{-0.03}^{+0.03}$ \\
J13283--023W & $3395_{-20}^{+27}$ & $+0.03_{-0.04}^{+0.05}$ \\
J13293+114 & $3407_{-19}^{+41}$ & $-0.11_{-0.04}^{+0.04}$ \\
J13299+102 & $3682_{-9}^{+15}$ & $-0.09_{-0.03}^{+0.02}$ \\
J13427+332 & $3154_{-11}^{+33}$ & $-0.04_{-0.04}^{+0.01}$ \\
J13450+176 & $3627_{-15}^{+20}$ & $-0.46_{-0.05}^{+0.06}$ \\
J13457+148$^{1,2}$ & $3618_{-11}^{+2}$ & $-0.24_{-...}^{+...}$ \\
J13458--179 & $3247_{-28}^{+37}$ & $+0.05_{-0.05}^{+0.03}$ \\
J13536+776 & $3367_{-36}^{+237}$ & $-0.06_{-0.06}^{+0.06}$ \\
J13582+125 & $3288_{-32}^{+42}$ & $-0.19_{-0.04}^{+0.04}$ \\
J14010--026 & $3707_{-11}^{+11}$ & $-0.09_{-0.04}^{+0.01}$ \\
J14082+805 & $3797_{-11}^{+5}$ & $+0.09_{-0.03}^{+0.04}$ \\
J14152+450 & $3414_{-27}^{+30}$ & $-0.00_{-0.05}^{+0.02}$ \\
J14173+454$^{3}$ & $3088_{-97}^{+345}$ & $+0.16_{-0.06}^{+0.08}$ \\
J14251+518$^{2}$ & $3419_{-12}^{+18}$ & $-0.09_{-0.05}^{+0.01}$ \\
J14257+236E & $3841_{-5}^{+6}$ & $+0.21_{-0.03}^{+0.05}$ \\
J14257+236W & $3875_{-7}^{+5}$ & $+0.25_{-0.04}^{+0.03}$ \\
J14294+155 & $3674_{-12}^{+14}$ & $+0.07_{-0.01}^{+0.05}$ \\
J14307--086 & $3905_{-10}^{+7}$ & $+0.14_{-0.02}^{+0.06}$ \\
J14310--122 & $3120_{-12}^{+51}$ & $+0.11_{-0.05}^{+0.03}$ \\
J14342--125 & $3056_{-33}^{+35}$ & $+0.15_{-0.06}^{+0.02}$ \\
J14524+123 & $3641_{-13}^{+11}$ & $+0.23_{-0.05}^{+0.02}$ \\
J14544+355 & $3309_{-22}^{+41}$ & $+0.01_{-0.03}^{+0.06}$ \\
J15013+055 & $3330_{-17}^{+31}$ & $-0.01_{-0.04}^{+0.01}$ \\
J15095+031 & $3451_{-10}^{+19}$ & $+0.02_{-0.04}^{+0.03}$ \\
J15194--077$^{2}$ & $3259_{-13}^{+31}$ & $-0.14_{-0.03}^{+0.00}$ \\
J15218+209 & $3651_{-15}^{+77}$ & $-0.16_{-0.09}^{+0.03}$ \\
J15369--141 & $3117_{-30}^{+49}$ & $+0.10_{-0.06}^{+0.03}$ \\
J15474--108 & $3483_{-28}^{+81}$ & $-0.37_{-0.05}^{+0.07}$ \\
J15499+796$^{3}$ & $3123_{-120}^{+335}$ & $+0.01_{-0.21}^{+0.09}$ \\
J15598--082 & $3688_{-8}^{+12}$ & $-0.05_{-0.02}^{+0.04}$ \\
J16028+205 & $3132_{-23}^{+47}$ & $+0.01_{-0.06}^{+0.03}$ \\
J16092+093 & $3402_{-16}^{+28}$ & $-0.08_{-0.03}^{+0.02}$ \\
J16167+672N & $3472_{-9}^{+18}$ & $+0.07_{-0.04}^{+0.03}$ \\
J16167+672S & $3926_{-17}^{+7}$ & $+0.16_{-0.02}^{+0.06}$ \\
J16254+543 & $3533_{-21}^{+10}$ & $-0.32_{-0.03}^{+0.02}$ \\
J16303--126 & $3150_{-11}^{+39}$ & $-0.08_{-0.02}^{+0.03}$ \\
J16327+126 & $3363_{-18}^{+36}$ & $-0.21_{-0.02}^{+0.03}$ \\
J16462+164 & $3427_{-10}^{+11}$ & $+0.02_{-0.01}^{+0.04}$ \\
J16554--083N & $3150_{-17}^{+33}$ & $-0.17_{-0.04}^{+0.03}$ \\
J16570--043$^{3}$ & $3335_{-41}^{+264}$ & $-0.05_{-0.09}^{+0.04}$ \\
J16581+257$^{2}$ & $3732_{-13}^{+8}$ & $-0.03_{-0.01}^{+0.05}$ \\
J17033+514 & $3027_{-25}^{+67}$ & $+0.25_{-0.06}^{+0.03}$ \\
J17052--050 & $3536_{-9}^{+14}$ & $-0.24_{-0.01}^{+0.01}$ \\
J17071+215 & $3379_{-12}^{+17}$ & $-0.05_{-0.04}^{+0.02}$ \\
J17115+384 & $3242_{-12}^{+34}$ & $-0.01_{-0.03}^{+0.03}$ \\
J17166+080 & $3486_{-10}^{+19}$ & $-0.06_{-0.04}^{+0.01}$ \\
J17198+417 & $3427_{-19}^{+11}$ & $-0.22_{-0.02}^{+0.02}$ \\
J17303+055 & $3738_{-10}^{+15}$ & $-0.12_{-0.04}^{+0.01}$ \\
J17355+616 & $3807_{-8}^{+20}$ & $-0.01_{-0.02}^{+0.04}$ \\
J17364+683 & $3392_{-12}^{+25}$ & $+0.03_{-0.04}^{+0.03}$ \\
J17378+185 & $3554_{-10}^{+28}$ & $-0.23_{-0.03}^{+0.01}$ \\
J17542+073 & $3115_{-26}^{+70}$ & $-0.12_{-0.02}^{+0.05}$ \\
J17578+046 & $3212_{-27}^{+75}$ & $-0.29_{-0.07}^{+0.06}$ \\
J17578+465 & $3395_{-19}^{+19}$ & $+0.06_{-0.02}^{+0.04}$ \\
J18027+375 & $3098_{-21}^{+123}$ & $+0.23_{-0.05}^{+0.05}$ \\
J18051--030 & $3656_{-9}^{+8}$ & $-0.16_{-0.02}^{+0.01}$ \\
J18165+048 & $3056_{-27}^{+78}$ & $+0.31_{-0.04}^{+0.02}$ \\
J18174+483 & $3656_{-27}^{+24}$ & $-0.01_{-0.06}^{+0.04}$ \\
J18180+387E & $3325_{-16}^{+23}$ & $-0.21_{-0.04}^{+0.02}$ \\
J18189+661$^{3}$ & $3055_{-58}^{+372}$ & $+0.28_{-0.11}^{+0.04}$ \\
J18198--019 & $3917_{-7}^{+10}$ & $-0.11_{-0.04}^{+0.02}$ \\
J18221+063 & $3276_{-19}^{+47}$ & $-0.23_{-0.03}^{+0.03}$ \\
J18224+620 & $3094_{-28}^{+84}$ & $+0.18_{-0.12}^{+0.04}$ \\
J18319+406 & $3328_{-15}^{+32}$ & $-0.03_{-0.03}^{+0.03}$ \\
J18346+401 & $3162_{-18}^{+59}$ & $+0.23_{-0.06}^{+0.03}$ \\
J18353+457 & $3835_{-7}^{+10}$ & $+0.02_{-0.03}^{+0.03}$ \\
J18363+136 & $3259_{-34}^{+105}$ & $+0.18_{-0.09}^{+0.04}$ \\
J18409--133 & $3746_{-8}^{+8}$ & $+0.09_{-0.01}^{+0.06}$ \\
J18419+318 & $3365_{-12}^{+26}$ & $-0.09_{-0.05}^{+0.01}$ \\
J18427+596N & $3318_{-15}^{+34}$ & $-0.24_{-0.02}^{+0.03}$ \\
J18427+596S & $3230_{-24}^{+37}$ & $-0.28_{-0.04}^{+0.03}$ \\
J18480--145 & $3371_{-13}^{+27}$ & $-0.14_{-0.05}^{+0.01}$ \\
J18482+076 & $3098_{-38}^{+258}$ & $+0.19_{-0.09}^{+0.06}$ \\
J18498--238 & $3294_{-59}^{+180}$ & $-0.08_{-0.07}^{+0.12}$ \\
J18580+059 & $3829_{-5}^{+14}$ & $+0.03_{-0.01}^{+0.06}$ \\
J19070+208 & $3444_{-17}^{+16}$ & $-0.32_{-0.02}^{+0.01}$ \\
J19072+208 & $3388_{-11}^{+17}$ & $-0.32_{-0.02}^{+0.02}$ \\
J19084+322 & $3327_{-20}^{+24}$ & $-0.09_{-0.03}^{+0.01}$ \\
J19098+176 & $3047_{-14}^{+67}$ & $+0.26_{-0.05}^{+0.02}$ \\
J19169+051N & $3595_{-12}^{+9}$ & $+0.05_{-0.01}^{+0.04}$ \\
J19216+208 & $3049_{-13}^{+82}$ & $+0.21_{-0.06}^{+0.02}$ \\
J19251+283 & $3270_{-20}^{+34}$ & $+0.05_{-0.04}^{+0.03}$ \\
J19346+045 & $3874_{-8}^{+13}$ & $-0.25_{-0.05}^{+0.02}$ \\
J19511+464$^{3}$ & $3284_{-56}^{+288}$ & $-0.36_{-0.11}^{+0.12}$ \\
J20260+585 & $3041_{-22}^{+101}$ & $+0.27_{-0.03}^{+0.03}$ \\
J20305+654 & $3386_{-22}^{+31}$ & $-0.07_{-0.04}^{+0.02}$ \\
J20336+617 & $3261_{-39}^{+36}$ & $+0.23_{-0.09}^{+0.05}$ \\
J20405+154 & $3041_{-12}^{+90}$ & $+0.26_{-0.04}^{+0.03}$ \\
J20450+444 & $3556_{-14}^{+8}$ & $-0.11_{-0.04}^{+0.01}$ \\
J20525--169 & $3135_{-15}^{+61}$ & $+0.17_{-0.06}^{+0.03}$ \\
J20533+621 & $3783_{-5}^{+11}$ & $+0.03_{-0.01}^{+0.05}$ \\
J20556--140S & $3040_{-23}^{+84}$ & $+0.04_{-0.08}^{+0.08}$ \\
J20567--104 & $3505_{-17}^{+8}$ & $+0.06_{-0.02}^{+0.04}$ \\
J21019--063 & $3500_{-13}^{+11}$ & $+0.05_{-0.02}^{+0.04}$ \\
J21152+257 & $3568_{-16}^{+36}$ & $+0.34_{-0.07}^{+0.02}$ \\
J21164+025 & $3373_{-10}^{+20}$ & $+0.04_{-0.03}^{+0.04}$ \\
J21221+229 & $3716_{-15}^{+10}$ & $-0.15_{-0.02}^{+0.02}$ \\
J21348+515 & $3494_{-14}^{+12}$ & $-0.08_{-0.03}^{+0.03}$ \\
J21463+382 & $3174_{-44}^{+32}$ & $-0.38_{-0.09}^{+0.04}$ \\
J21466+668 & $3134_{-15}^{+55}$ & $+0.02_{-0.04}^{+0.03}$ \\
J21466--001 & $3130_{-16}^{+35}$ & $+0.02_{-0.06}^{+0.02}$ \\
J22012+283$^{3}$ & $3559_{-216}^{+145}$ & $-0.58_{-0.29}^{+0.16}$ \\
J22020--194 & $3207_{-11}^{+49}$ & $-0.05_{-0.02}^{+0.03}$ \\
J22021+014 & $3839_{-15}^{+8}$ & $-0.01_{-0.01}^{+0.05}$ \\
J22057+656 & $3664_{-20}^{+8}$ & $-0.11_{-0.02}^{+0.03}$ \\
J22096--046 & $3372_{-18}^{+29}$ & $+0.17_{-0.04}^{+0.05}$ \\
J22115+184 & $3673_{-5}^{+12}$ & $+0.21_{-0.03}^{+0.03}$ \\
J22125+085 & $3370_{-10}^{+12}$ & $-0.11_{-0.03}^{+0.01}$ \\
J22137--176 & $3056_{-15}^{+78}$ & $+0.09_{-0.07}^{+0.06}$ \\
J22252+594 & $3236_{-28}^{+48}$ & $+0.16_{-0.08}^{+0.03}$ \\
J22298+414 & $3115_{-27}^{+56}$ & $-0.00_{-0.06}^{+0.04}$ \\
J22330+093 & $3605_{-9}^{+25}$ & $-0.15_{-0.04}^{+0.01}$ \\
J22503--070 & $3824_{-8}^{+6}$ & $-0.09_{-0.02}^{+0.01}$ \\
J22518+317$^{3}$ & $3469_{-44}^{+191}$ & $-0.22_{-0.11}^{+0.03}$ \\
J22532--142 & $3125_{-14}^{+43}$ & $+0.11_{-0.05}^{+0.02}$ \\
J22559+178 & $3771_{-11}^{+13}$ & $+0.03_{-0.02}^{+0.05}$ \\
J22565+165$^{1}$ & $3713_{-3}^{+20}$ & $+0.11_{-0.01}^{+0.06}$ \\
J23113+085 & $3268_{-26}^{+33}$ & $+0.09_{-0.05}^{+0.02}$ \\
J23216+172 & $3184_{-20}^{+61}$ & $+0.12_{-0.05}^{+0.08}$ \\
J23245+578 & $3786_{-8}^{+6}$ & $+0.15_{-0.02}^{+0.05}$ \\
J23340+001 & $3501_{-10}^{+20}$ & $-0.15_{-0.03}^{+0.02}$ \\
J23381--162 & $3481_{-11}^{+14}$ & $-0.20_{-0.02}^{+0.01}$ \\
J23419+441$^{1,2}$ & $3136_{-4}^{+208}$ & $+0.23_{-0.02}^{+0.01}$ \\
J23431+365 & $3131_{-22}^{+113}$ & $+0.21_{-0.08}^{+0.04}$ \\
J23492+024 & $3447_{-10}^{+26}$ & $-0.40_{-0.03}^{+0.02}$ \\
J23505--095 & $3095_{-15}^{+37}$ & $+0.20_{-0.06}^{+0.02}$ \\
J23556--061 & $3624_{-15}^{+18}$ & $+0.23_{-0.06}^{+0.01}$ \\
J23585+076 & $3487_{-32}^{+42}$ & $-0.05_{-0.03}^{+0.03}$ \\
\end{xtabular}
\end{center}
\tablefoot{
    \tablefoottext{a}{Stars marked with ``1'' and ``2'' were used for training the DTL $T_{\rm eff}$ and [M/H] models, respectively. The number ``3'' refers to stars with rotational velocity ${\varv\sin{i} > \mathrm{10\,km\,s^{-1}}}$.
    Some $1 \sigma$ uncertainties and quantiles for stars marked 1 and 2 are missing because the different predictions were too close. In these few cases, the quantiles could be substituted by the standard deviation.}
}

\end{appendix}
\end{document}